\documentclass[a4paper,preprint]{aastex}

\usepackage{natbib}

\shorttitle{TrES-2, TrES-3, WASP-3 and HAT-P-4}
\shortauthors{Christiansen et al.}

\begin{document}

\title{System parameters, transit times and secondary eclipse constraints of the exoplanet systems HAT-P-4, TrES-2, TrES-3 and WASP-3 from the NASA {\it EPOXI} Mission of Opportunity.}

\author{Jessie~L.~Christiansen\altaffilmark{1}, Sarah~Ballard\altaffilmark{1}, David~Charbonneau\altaffilmark{1}, Drake~Deming\altaffilmark{2}, Matthew~J.~Holman\altaffilmark{1}, Nikku~Madhusudhan\altaffilmark{3}, Sara~Seager\altaffilmark{3}, Dennis~D.~Wellnitz\altaffilmark{4}, Richard~K.~Barry\altaffilmark{2}, Timothy~A.~Livengood\altaffilmark{2}, Tilak~Hewagama\altaffilmark{2,4}, Don~L.~Hampton\altaffilmark{5}, Carey~M.~Lisse\altaffilmark{6}, and Michael~F.~A'Hearn\altaffilmark{4}}

\altaffiltext{1}{Harvard-Smithsonian Center for Astrophysics, 60 Garden Street, Cambridge, MA 02138, USA; jchristi@cfa.harvard.edu}
\altaffiltext{2}{Goddard Space Flight Center, Greenbelt, MD 20771, USA}
\altaffiltext{3}{Massachusetts Institute of Technology, Cambridge, MA 02159, USA}
\altaffiltext{4}{University of Maryland, College Park, MD 20742, USA}
\altaffiltext{5}{University of Alaska Fairbanks, Fairbanks AK 99775, USA}
\altaffiltext{6}{Johns Hopkins University Applied Physics Laboratory, Laurel, MD 20723, USA}


\begin{abstract}
As part of the NASA {\it EPOXI} Mission of Opportunity, we observed seven known transiting extrasolar planet systems in order to construct time series photometry of extremely high phase coverage and precision. Here we present the results for four ``hot-Jupiter systems'' with near-solar stars---HAT-P-4, TrES-3, TrES-2 and WASP-3. We observe ten transits of HAT-P-4, estimating the planet radius $R_p=1.332\pm0.052$ $R_{\rm Jup}$, the stellar radius $R_{\star} = 1.602\pm0.061$ $R_{\odot}$, the inclination $i = 89.67\pm0.30$ degrees and the transit duration from first to fourth contact $\tau=255.6\pm1.9$ minutes. For TrES-3, we observe seven transits, and find $R_p=1.320\pm0.057$ $R_{\rm Jup}$, $R_{\star}=0.817\pm0.022$ $R_{\odot}$, $i=81.99\pm0.30$ degrees and $\tau=81.9\pm1.1$ minutes. We also note a long term variability in the TrES-3 light curve, which may be due to star spots. We observe nine transits of TrES-2, and find $R_p=1.169\pm0.034$ $R_{\rm Jup}$, $R_{\star}=0.940\pm0.026$ $R_{\odot}$, $i=84.15\pm0.16$ degrees and $\tau=107.3\pm1.1$ minutes. Finally we observe eight transits of WASP-3, finding $R_p=1.385\pm0.060$ $R_{\rm Jup}$, $R_{\star}=1.354\pm0.056$ $R_{\odot}$, $i=84.22\pm0.81$ degrees and $\tau=167.3\pm1.3$ minutes. We present refined orbital periods and times of transit for each target. We state 95\% confidence upper limits on the secondary eclipse depths in our broadband visible bandpass centered on 650 nm. These limits are 0.073\% for HAT-P-4, 0.062\% for TrES-3, 0.16\% for TrES-2 and 0.11\% for WASP-3. We combine the TrES-3 secondary eclipse information with the existing published data and confirm that the atmosphere likely does not have a temperature inversion.

\end{abstract}


\section{Introduction}
\label{sec:intro}

The EPOXI Mission of Opportunity is a re-purposing of the Deep Impact flyby spacecraft, and comprises the Extrasolar Planet Observation and Characterization (EPOCh) investigation and the Deep Impact eXtended Investigation (DIXI). The primary goal of EPOCh was to scrutinize a small set of known transiting extrasolar planets. From 2008 January to 2008 August, we used the high resolution imaging (HRI) instrument \citep{Hampton05} and a broadband visible filter to construct high precision, high phase coverage and high cadence light curves for seven targets. We observed each target nearly continuously for several weeks at a time. The main science goals of EPOCh were to refine the system parameters of the known planets, to search for additional planets both directly (via transits of the additional body) and indirectly (via induced changes in the transits of the known planet), and to constrain the reflected light from the known planet at secondary eclipse. It is also useful to provide updated periods and times of epoch for these systems in order to reduce uncertainties on predicted transit and eclipse times, and therefore maximize the return of follow-up observations. In previous EPOCh papers we have presented the search for additional planets in the GJ~436 system \citep{Ballard10} and the secondary eclipse constraints for HAT-P-7 \citep{Christiansen10}. In this paper we present the updated system parameters, including constraints on the transit timing and changes in the transit parameters, and secondary eclipse constraints for a further four targets: HAT-P-4, TrES-3, TrES-2 and WASP-3, introduced below. The search for additional planets in these systems will be presented in a separate paper (Ballard et al. in prep). 

The exoplanet HAT-P-4b \citep{Kovacs07}, orbits a slightly evolved metal-rich late F star. With a mass of 0.68 $M_{\rm Jup}$ and a radius of 1.27 $R_{\rm Jup}$, it joined the ranks of inflated planets that have continued to challenge models of the physical structure of hot Jupiters.

TrES-3 \citep{ODonovan07} is notable for its very short orbital period of 1.30619 days. This proximity to the star makes TrES-3 a promising target for observations of reflected light at visible wavelengths; the planet-to-star flux ratio as measured in reflected light during the secondary eclipse is given by $A_g(R_p/a)^2$, where $A_g$ is the geometric albedo, $R_p$ is the planetary radius and $a$ is the semi-major axis of the planetary orbit. \citet{Winn08}, \citet{deMooij09} and \citet{Fressin10} have observed secondary eclipses of TrES-3 at visible and near-infrared wavelengths, and the emerging picture of the planetary atmosphere is one with efficient day-night re-circularization and no temperature inversion in the upper atmosphere. This is in contrast to predictions of a temperature inversion based on the high level of irradiation \citep{Fortney08}. \citet{Sozzetti09} studied the transit timing variations of TrES-3 and noted significant outliers from a constant period. \citet{Gibson09} monitored further transit times of TrES-3 and ruled out sub-Earth mass planets in the exterior and interior 2:1 resonances for circular orbits.

TrES-2 \citep{ODonovan06} was the first transiting planet found in the field of view of the NASA {\it Kepler} mission \citep{Borucki09}. \citet{Holman07} noted that the high impact parameter ($b \approx 0.85$) of TrES-2 made transit parameters such as inclination and duration sensitive to changes due to orbital precession. \citet{Mislis09} and \citet{Mislis10} claimed a significantly shorter duration for TrES-2 transits two years after the measurements of \citet{Holman07}. They proposed that this was caused by a change in orbit inclination due to precession, and that the duration would continue to decrease. However, \citet{Scuderi09} measured a duration consistent with \citet{ODonovan06} and \citet{Holman07} and did not see the predicted trend of decreasing transit duration. Secondary eclipses of TrES-2 have been observed in the near-infrared \citep{ODonovan10}, and the results favor a thermal inversion in the upper atmosphere, supporting the hypothesis that highly irradiated planetary atmospheres have inversions. The transit timing variations of TrES-2 have been studied by \citet{Raetz09} and \citet{Rabus09}, who find no statistically significant variations.

WASP-3b \citep{Pollacco08}, with a short period (1.84634 days) and a hot host star (F7-F8V, $T_{\rm eff}=6400$K), is one of the hottest transiting planets known, and another very good target for observing reflected light at secondary eclipse.

The paper is organized as follows. The observations and generation of the light curves are described in Section \ref{sec:data}, the transit analysis is presented in Section \ref{sec:res}, the secondary eclipse analysis is presented in Section \ref{sec:eclipses} and the results are discussed in Section \ref{sec:disc}.

\section{Observations and analysis}
\label{sec:data}

The EPOCh observations were made using the high resolution imager (HRI), which has a 30-cm aperture and a 1024$\times$1024 pixel CCD. For our observations we used a clear visible filter, covering 350--1000nm, in order to maximize the throughput of photons. The integration time for the science observations was 50 seconds, which for near-continuous observations results in roughly 1500 images per day. Since the on-board spacecraft memory is only 300Mb, we initially chose to read out only a $128\times128$ pixel sub-array of the full CCD, to ensure full phase coverage between data downlinks from the spacecraft. The CCD comprises four quadrants that are read out independently, and the sub-array is centered on the CCD where the four quadrants meet. The pixel scale is 0.4 arcsec pixel$^{-1}$, resulting in a sub-array field of view of 0.72 square arcminutes. The images are significant defocused, resulting in a stellar point-spread function (PSF) with a full-width half maximum of 4 arcseconds. Typically this meant that the target star was the only star in the field of view, and we were unable to employ relative photometry techniques for removing correlated noise in the light curves.

Table \ref{tab:obs} summarizes the observing schedules for each of the four targets. HAT-P-4 and TrES-3, along with GJ436 and XO-2, were observed during the initial observing block from 2008 January to 2008 May. The project was awarded an additional contingent observing block from 2008 June to 2008 August, during which time HAT-P-4 was re-observed, and TrES-2 and WASP-3 were also observed. During the contingent observations we began observing in a larger $256\times256$ pixel sub-array mode, to reduce losses from pointing drifts that occasionally resulted in the target star lying outside of the $128\times128$ pixel sub-array field of view. The number of images that could be obtained with the larger sub-array mode between data downlinks from the spacecraft was constrained by the data storage capacity on board the spacecraft. Therefore, in order to maximize the phase coverage we chose to restrict observations in the $256\times256$ pixel sub-array mode to the times of particular interest---during the transits and secondary eclipses. One event per data downlink could be observed in the larger sub-array mode without reducing the temporal coverage. Table \ref{tab:obs} shows the total number of transits and eclipses observed for each target, with the number observed in the $256\times256$ pixel sub-array mode given in parentheses. As discussed in Section \ref{sec:intro}, TrES-2 was claimed to show changes in the transit inclination with time. Therefore, we used the larger sub-array mode to observe the transits of TrES-2 where possible. WASP-3 was a promising target for secondary eclipse observations, and therefore we observed the secondary eclipses of WASP-3 in the larger mode where possible. For HAT-P-4 we observed two of the three transits and two of the three eclipses obtained in the contingent observations in the $256\times256$ pixel sub-array mode. TrES-3 was observed in the initial observing block and no observations were obtained in the larger mode.

\begin{deluxetable}{lclcc}
\tabletypesize{\scriptsize}
\tablecaption{EPOCh observations}
\tablewidth{0pt}
\tablehead{\colhead{Target} & \colhead{{\it V} Mag} & \colhead{UT Dates observed (2008)} & \colhead{No. of Transits\tablenotemark{a}} & \colhead{No. of Eclipses\tablenotemark{a}}}
\startdata
HAT-P-4 & 11.22 & 01/22--02/12, 06/29--07/07               & 10 (2) & 9 (2) \\
TrES-3  & 11.18 & 03/06--03/18                             & 7 (0)  & 6 (0) \\
TrES-2  & 11.41 & 06/27--06/28, 07/19--07/29               & 9 (7)  & 8 (2) \\
WASP-3  & 10.64 & 07/17--07/18, 07/30--08/07, 08/10--08/15 & 8 (0)  & 9 (8) \\
\enddata
\tablenotetext{a}{Including partial events. The number in brackets is the subset of events observed in $256\times256$ pixel sub-array mode.}
\label{tab:obs}
\end{deluxetable}

\subsection{Image Calibration and Time Series Extraction}
\label{sec:timeseries}

We receive calibrated FITS images from the extant Deep Impact data reduction pipeline \citep{Klaasen05}. These data have been bias- and dark-subtracted and flat-fielded, using calibration images obtained on the ground before launch. Due to the very high precision required in the light curves, we perform several additional calibration steps to account for changes in the CCD since launch. The spacecraft pointing drifts considerably with time, resulting in significant coverage of the CCD by the stellar PSF and placing paramount importance on the flat-fielding. The procedure is described in \citet{Ballard10} and summarized here.

For each target, we use a PSF constructed from the images to locate the position of the star to a hundredth of a pixel. At this stage we reject images with $10\sigma$ outliers from the PSF fit, assuming the stellar PSF to be contaminated by an energetic particle hit. We subtract a time-dependent bias calculated for each quadrant from the corresponding overscan region. We reduce the pixels in the central columns and rows of the CCD (forming the internal boundaries between the quadrants) by roughly 15\% and 1\% respectively, to correct an artifact produced by the CCD readout electronics. For data obtained in the $256\times256$ pixel mode, we scale the images by a constant (typically differing from unity by one part in a thousand) to correct an observed flux offset between the two sub-array modes.

In order to track time-dependent changes in the flat-field since launch, there is a small green LED stimulation lamp that can be switched on to illuminate the CCD. We obtained blocks of 200 calibration frames using this lamp, which were taken every few days throughout the observations, alternating between blocks in the smaller and larger sub-array modes in the contingent observations. We correct each science frame by the flat-field generated from lamp images taken in the same sub-array mode. We assume any remaining flat-field errors to be color-dependent and therefore unable to be addressed by the monochromatic lamp.

We perform aperture photometry, using a circular aperture of radius 10 pixels. The resulting light curves exhibit significant correlated noise on the order of 1\%, which is associated with the drift in the spacecraft pointing. In order to correct for this, we use the data itself to generate a sensitivity map of the CCD. We assume the out-of-transit and out-of-eclipse data to be of uniform brightness, with two caveats. First, the star may have intrinsic variations in stellar brightness due to spots. Only one of the four targets displayed long-period variability (Figure \ref{fig:tres3_fulllc_noise}), and this was removed by fitting and removing a polynomial in time before producing the CCD sensitivity map. Second, transits of additional planets may be present, which will be suppressed with this treatment \citep{Ballard10}. We randomly draw several thousand of the out-of-transit and out-of-eclipse points and find a robust average flux of the 30 spatially nearest neighbors. We use this set of averages to generate a two-dimensional surface spline to the flux distribution across the CCD. Each point in the light curve is then corrected by interpolating onto this surface. The entire procedure is iterated several times to converge on the positions and scaling factors that result in the lowest scatter in the out-of-transit and out-of-eclipse data in the final light curve.

\begin{figure}[h!]
\begin{center}
 \includegraphics[width=3in, angle=90]{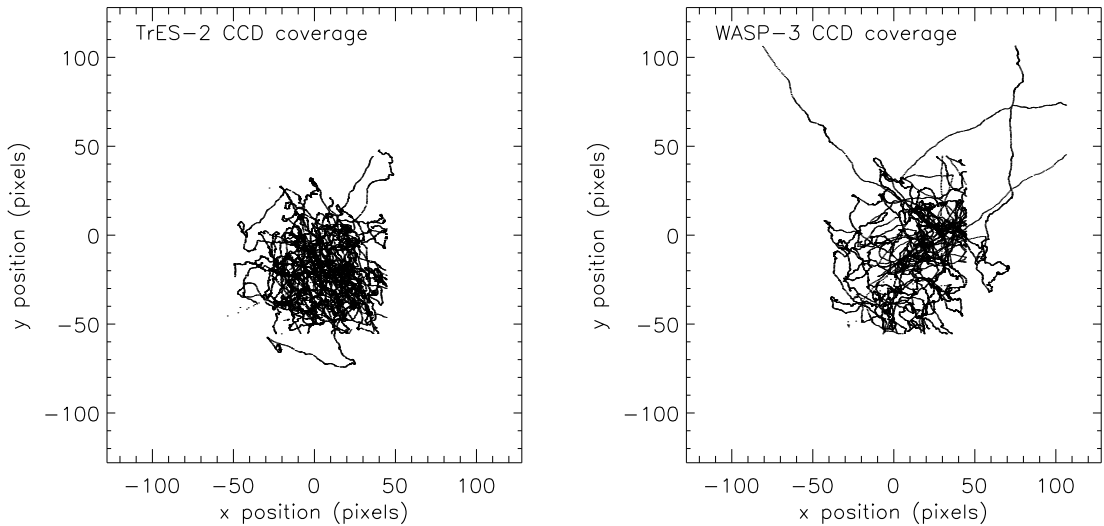}
 \caption{The CCD positions of the photometry apertures for two targets. {\it Left}: TrES-2 is confined to the center of the CCD and therefore the same pixels are sampled well for creating a robust surface spline. {\it Right}: WASP-3 is spread over a much larger fraction of the CCD, including large excursions out of the central 128$\times$128 pixel sub-array when observations were obtained in the 256$\times$256 pixel sub-array. This reduces the quality of the surface spline and results in a larger component of correlated noise in the WASP-3 light curve.}
   \label{fig:positions_compare}
\end{center}
\end{figure}

The robustness of the surface spline for each target depends on the coverage of the CCD by that target. If the coverage is small and the corresponding density of photometry apertures high, then there is a high probability that the same pixel will be returned to multiple times over the observations. Having flux measurements separated in time reduces the influence of stellar activity on our calibration of the sensitivity of each pixel. Figure \ref{fig:positions_compare} shows the complete CCD coverage for two targets. TrES-2 is well confined on the CCD and the density of photometry apertures leads to a more robust surface spline. The TrES-2 light curve prior to and post the application of the surface spline is shown in Figure \ref{fig:tres2_fulllc_noise}. On the other hand, the photometry apertures for WASP-3 sample a much larger area of the CCD, and in addition many of the observations obtained in the 256$\times$256 pixel sub-array mode do not overlay the central 128$\times$128 pixel sub-array. The resulting surface spline is therefore more sensitive to noise introduced by stellar activity or systematics that are not an artifact of the pointing jitter. The WASP-3 light curve prior to and post the application of the surface spline is shown in Figure \ref{fig:wasp3_fulllc_noise}. The lower panel of Figure \ref{fig:wasp3_fulllc_noise} shows how the noise in the final calibrated WASP-3 light curve bins down compared with the expectation for Gaussian noise, and the poor quality of the data is due to the low density of the CCD coverage for WASP-3.

\subsection{Details for each target}
\label{sec:details}

\begin{figure}[h!]
\begin{center}
 \includegraphics[width=4in, angle=90]{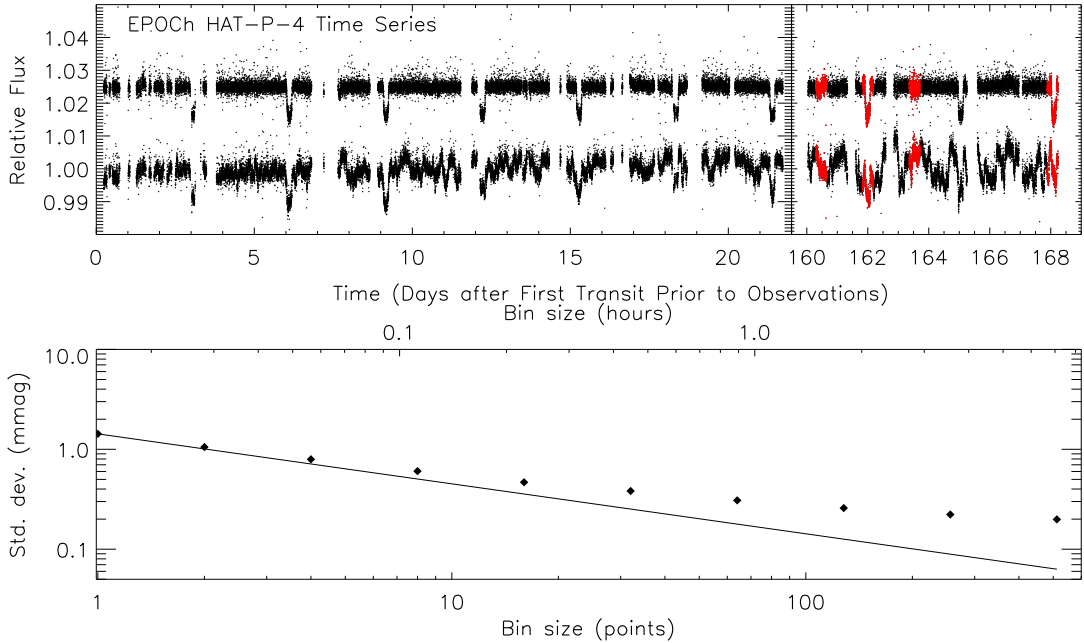}
 \caption{{\it Upper panel}: The full HAT-P-4 EPOXI light curve. The left panel shows the original run of seven consecutive transits. The right panel shows the three transits observed five months later during the EPOCh contingent observations. In each panel the lower curve is before the first application of the surface spline and the upper curve is the final calibrated light curve. The red data points were obtained in the larger $256\times256$ pixel sub-array mode. {\it Lower panel}: The scatter in the out-of-transit data with increasing bin size (diamonds) and comparing to the expectation for Gaussian noise ($1/\sqrt{N}$, where $N$ is the number of points in the bin, shown as the solid line normalized to the unbinned value of the scatter). The points do not follow the line, indicating residual correlated noise in the light curve.}
   \label{fig:hatp4_fulllc_noise}
\end{center}
\end{figure}

The final HAT-P-4 light curve is shown in Figure \ref{fig:hatp4_fulllc_noise}. HAT-P-4 was the first EPOCh target observed, initially for 22 days from 2008 January 22 to 2008 February 12, during the original EPOCh target schedule, and again for 8 days from 2008 June 29 to 2008 July 7 during the contingent observations. Of the 45,320 images obtained of HAT-P-4, 5434 were discarded due to the star being either out of the field of view or too close to the edge of the CCD to measure accurate photometry, 1305 were discarded due to energetic particle hits, and 76 were discarded due to readout smear, for a final total of 38,505 acceptable images. All of the data obtained in the initial run are in the $128\times128$ pixel sub-array mode. Of the contingent data, two of the three transits and two of the three eclipses are in the larger $256\times256$ pixel sub-array mode, and the remaining data are in the smaller mode. The bottom panel of Figure \ref{fig:hatp4_fulllc_noise} shows how the scatter in the final light curve scales down with increasing bin size---for Gaussian noise the expectation is the scatter will decrease as $1/\sqrt{N}$, where $N$ is the number of points in the bin.

\begin{figure}[h!]
\begin{center}
 \includegraphics[width=4in, angle=90]{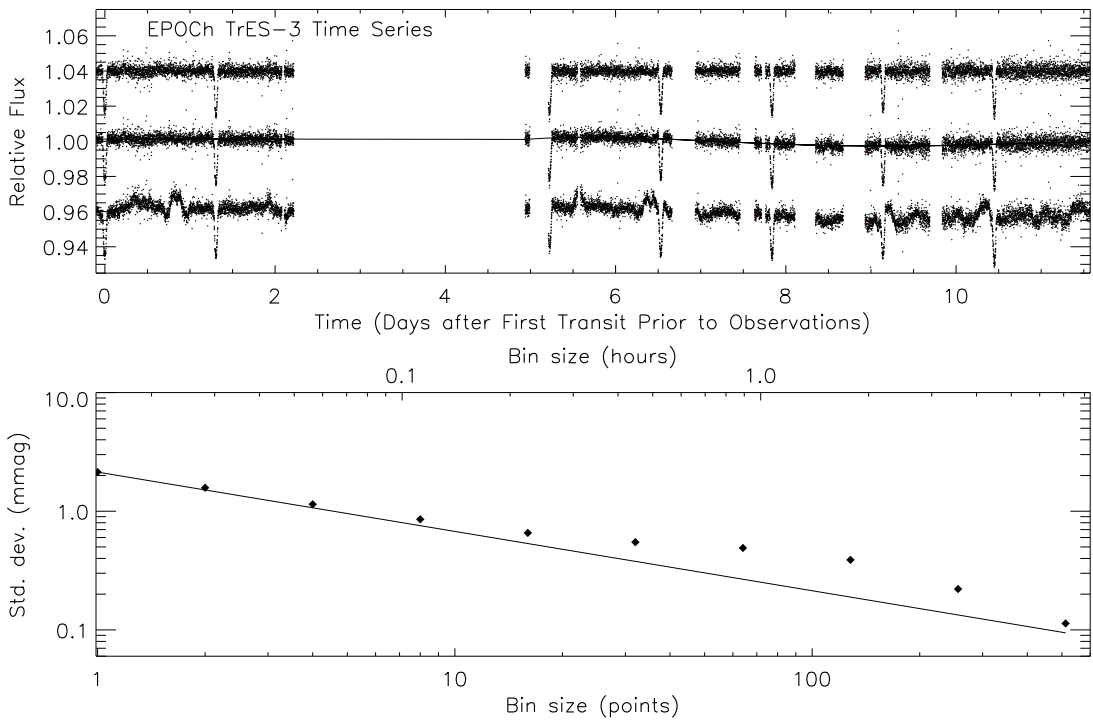}
 \caption{{\it Upper panel}: The TrES-3 EPOXI light curve. The gap from 2.5--5 days is during the pre-look for a subsequent target. Seven transits of TrES-3 were observed in total. The lowest light curve is prior to the first application of the surface spline, the middle light curve is after the application of the spline but prior to the removal of the time-dependent polynomial, and the upper light curve is the final calibrated data set. {\it Lower panel}: See Figure \ref{fig:hatp4_fulllc_noise} for explanation. In the case of TrES-3, where all data were obtained in the smaller sub-array mode and the total time span is relatively short, the scatter bins down close to the expectation for Gaussian noise.}
   \label{fig:tres3_fulllc_noise}
\end{center}
\end{figure}

The final TrES-3 light curve is shown in Figure \ref{fig:tres3_fulllc_noise}. TrES-3 was the second EPOCh target observed, for 12 days from 2008 March 6 to 2008 March 18. The gap in the light curve from 2--5 days is due to a `pre-look' for the subsequent EPOCh target, XO-2, which was performed in order to refine the pointing for that target. We obtained a total of 14,195 images of TrES-3, of which we discarded 1165 due to the star being out of or too close to the edge of the field of view, 1632 due to energetic particle hits and 127 due to readout smear, leaving 11,271 images. We obtained all of the TrES-3 data in the $128\times128$ pixel sub-array mode. After the initial application of the two-dimensional surface spline a long timescale, low amplitude variability was evident in the light curve. This can be seen in the lower light curve in Figure \ref{fig:tres3_fulllc_noise}. In order to remove this variability we bin the out-of-transit data by two hours and fit with a time-dependent fifth-order polynomial for the data occurring later than 4.0 days. We divide out this feature before iterating over the previous steps to produce the final light curve. The polynomial is plotted on the lower light curve, and the final light curve is shown as the upper curve in Figure \ref{fig:tres3_fulllc_noise}. As with HAT-P-4, the bottom panel of Figure \ref{fig:tres3_fulllc_noise} shows the noise properties of the data.

\begin{figure}[h!]
\begin{center}
 \includegraphics[width=4in, angle=90]{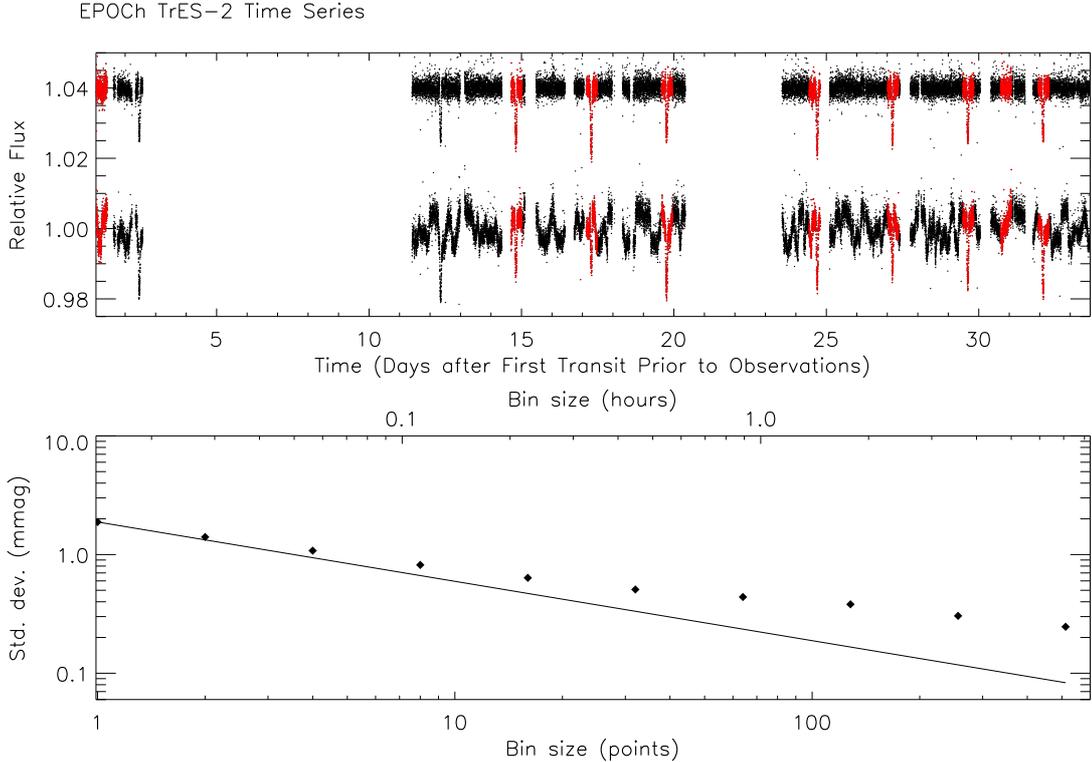}
 \caption{{\it Upper panel}: The TrES-2 EPOXI light curve. The data obtained from days 1--3 are the pre-look, for refinement of the spacecraft pointing. From days 3--11 the spacecraft was observing a different target before returning to TrES-2 with updated pointing parameters. The gap from days 21--23 spans the pre-look for the subsequent target. Nine transits of TrES-2 were observed in total. The lower curve is prior to the first application of the surface spline and the upper curve is the final calibrated light curve. The red data points were obtained in the larger $256\times256$ pixel sub-array mode. {\it Lower panel}: See Figure \ref{fig:hatp4_fulllc_noise} for explanation.}
   \label{fig:tres2_fulllc_noise}
\end{center}
\end{figure}

The final TrES-2 light curve is shown in Figure \ref{fig:tres2_fulllc_noise}. We observed TrES-2 during the contingent EPOCh observations, from 2008 July 7 to 2008 July 30, in addition to a pre-look for pointing on 2008 June 28 and 29. In total, we obtained 31,210 images of TrES-2, with 1979 discarded due to the star lying out of or too close to the edge of the field of view, 1427 discarded due to energetic particle hits and 80 discarded due to readout smear, for a total of 27,724 acceptable images. We observed nine transits in total, including seven in the $256\times256$ pixel sub-array mode. The lower panel of Figure \ref{fig:tres2_fulllc_noise} shows that correlated noise remains in the final light curve.

\begin{figure}[h!]
\begin{center}
 \includegraphics[width=4in, angle=90]{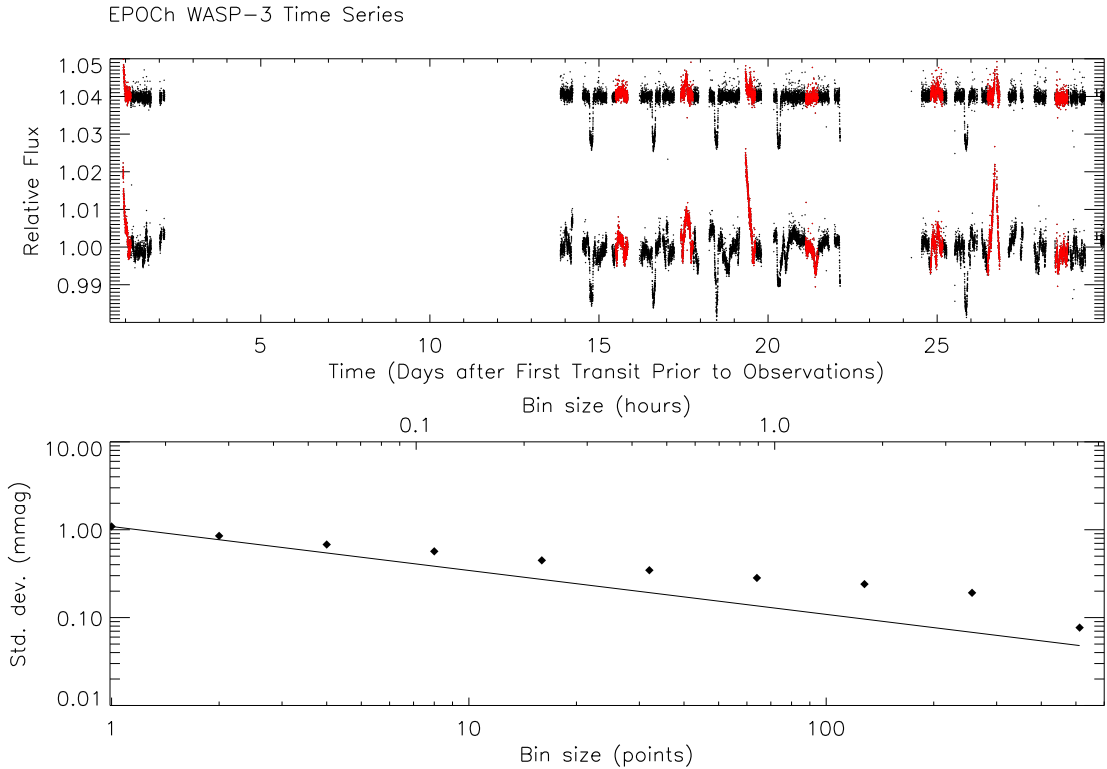}
 \caption{{\it Upper panel}: The WASP-3 EPOXI light curve. The first two days of data are the pre-look to refine the pointing. The gap between 22 and 24 days is due to the pre-look for the subsequent target. The significant positive deviations seen at 1.0, 17.6, 19.3 and 26.6 days are instrumental in nature; see the text for details. The lower curve is prior to the first application of the spatial spline and the upper curve is the final calibrated light curve. The red data points were obtained in the larger $256\times256$ pixel sub-array mode. {\it Lower panel}: See Figure \ref{fig:hatp4_fulllc_noise} for explanation.}
   \label{fig:wasp3_fulllc_noise}
\end{center}
\end{figure}

The final EPOCh light curve for WASP-3 is shown in Figure \ref{fig:wasp3_fulllc_noise}. We observed WASP-3 during the contingent observations, from 2008 July 29 to 2008 August 16, with a pre-look from 2008 July 17 to 2008 July 19. We obtained 24,015 images of WASP-3, of which we discarded 4,182 due to the star being out of or too close to the edge of the field of view, 403 due to energetic particle hits, and 808 due to readout smear, leaving 18,622 acceptable images. For WASP-3, none of the eight transits were observed in $256\times256$ pixel sub-array mode, however eight of the nine secondary eclipses were observed in this mode. The two-dimensional surface spline relies on multiple visits to the same part of the CCD to characterize robustly the interpixel variations. This is particularly true for the data that occur during the transits and eclipses, since they cannot be assumed to be of uniform flux and are therefore excluded from the creation of the surface. In order to effectively flat-field the data that are taken during transit and eclipse, the observations taken during these times must be gathered at the same spatial positions as data obtained at other times. In the case of WASP-3, four of the eight secondary eclipses occurred at locations that were poorly sampled. No out-of-transit or out-of-eclipse observations fell on these pixels, and therefore we cannot estimate the true sensitivity of these pixels in order to produce an effective flat-field. These eclipses occur at 1.0, 17.6, 19.3 and 26.6 days, and can be seen in the light curve as increases in flux. These four eclipses are discarded for the final analysis. Besides these events, a significant fraction of the WASP-3 data are distributed in poorly-sampled areas of the CCD, degrading the robustness of the two-dimensional surface spline. The bottom panel of Figure \ref{fig:wasp3_fulllc_noise} demonstrates the adverse effect this has on the noise properties of the final light curve, as the data do not bin down as expected for Gaussian noise.

\section{Transit analysis}
\label{sec:res}

For the transit analysis, we make several additional calibration steps. The two-dimensional surface spline uses only a fraction of the data to generate the surface, in order to preserve as much of the information in the light curve as possible, and to minimize the suppression of transits of putative additional planets. However for the transit analysis, we use all of the available data to calibrate each event. For each transit, we define a window approximately three times the duration of the transit, centered on the predicted transit time. We take each point in this window and divide the flux by a robust average of the 30 spatially nearest points that do not fall in any of the transit windows. This is essentially a point-by-point application of the full two-dimensional surface spline. We then fit a slope, linear with time, to the out-of-transit data across each transit and divide it out, to remove any residual long timescale trends.

For TrES-3, TrES-2 and WASP-3, we generate non-linear limb-darkening coefficients of the form given by \citet{Claret00}, $I_{\mu}/I_1 = 1 - \sum_{n=1}^4 c_n(1-\mu^{n/2})$, where $I_1$ is the specific intensity at the center of the disk and $\mu=cos(\gamma)$, with $\gamma$ the angle between the emergent intensity and the line of sight. We use photon-weighted stellar atmosphere models of \citet{Kurucz94,Kurucz05} that bracket the published values of stellar $T_{\rm eff}$ and log $g$, and convolve these with the total EPOXI response function, including filter, optics and CCD response. We fit for the four coefficients of the non-linear form of the limb-darkening using 17 positions across the stellar limb, at 2~nm intervals along the 350--1000~nm bandpass. We calculate the final set of coefficients as the weighted average when integrated over the bandpass, and bi-linearly interpolate across $T_{\rm eff}$ and log~$g$ for each target. The final set of coefficient for each targets is given in Table \ref{tab:tres3} for TrES-3, Table \ref{tab:tres2} for TrES-2 and Table \ref{tab:wasp3} for WASP-3. 


The quality of the EPOCh light curves is nearly sufficient to fit for the limb-darkening coefficients rather than assuming theoretical values. Ultimately, the degeneracies between the geometric parameters of the transiting system and the limb-darkening coefficients prevent us from placing meaningful constraints on the coefficients. In the case of HAT-P-4 however, the system is very close to edge-on ($i=89.9^{+0.1}_{-2.2}$ degrees), which reduces the parameter space considerably. Therefore, for HAT-P-4 we instead use a quadratic equation for the limb-darkening, $I_{\mu}/I_1 = 1 - a(1-\mu) - b(1-\mu)^2$, and allow two linear combinations of the coefficients, $c_1 = 2a + b$ and $c_2 = a-2b$ to be free parameters in the transit analysis, which produced a better fit to the data as defined below.

\begin{figure}[h!]
\begin{center}
 \includegraphics[width=4in, angle=90]{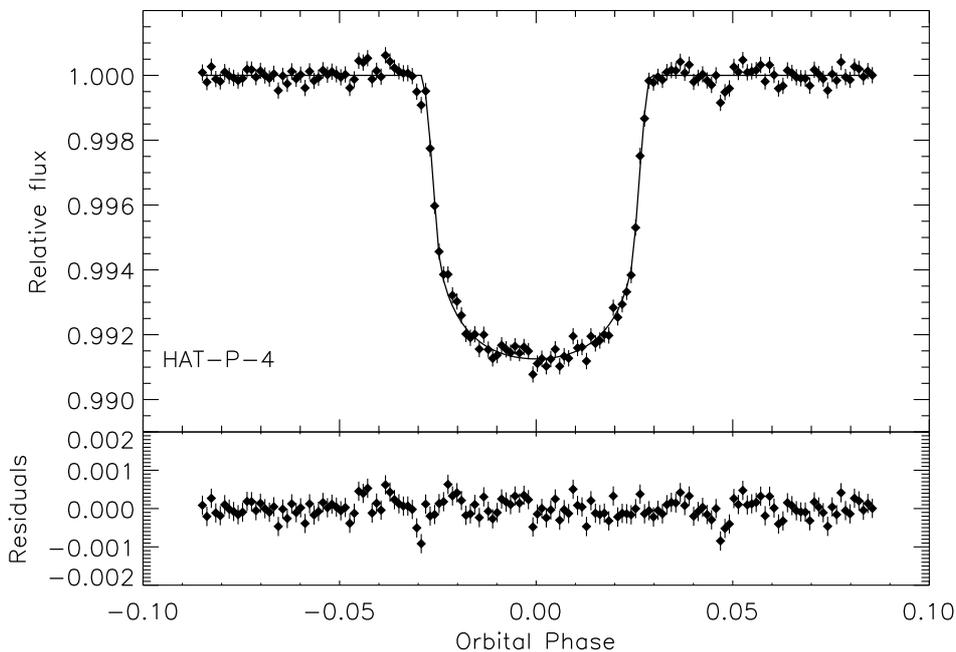}
 \caption{The seven HAT-P-4 transits from the original observing schedule, phase-folded and binned in five minute intervals. light curve. The solid line is the best fit transit model. {\it Lower panel}: The residuals when the best-fit model is subtracted from the data.}
   \label{fig:hatp4_phased}
\end{center}
\end{figure}

\begin{figure}[h!]
\begin{center}
 \includegraphics[width=4in, angle=90]{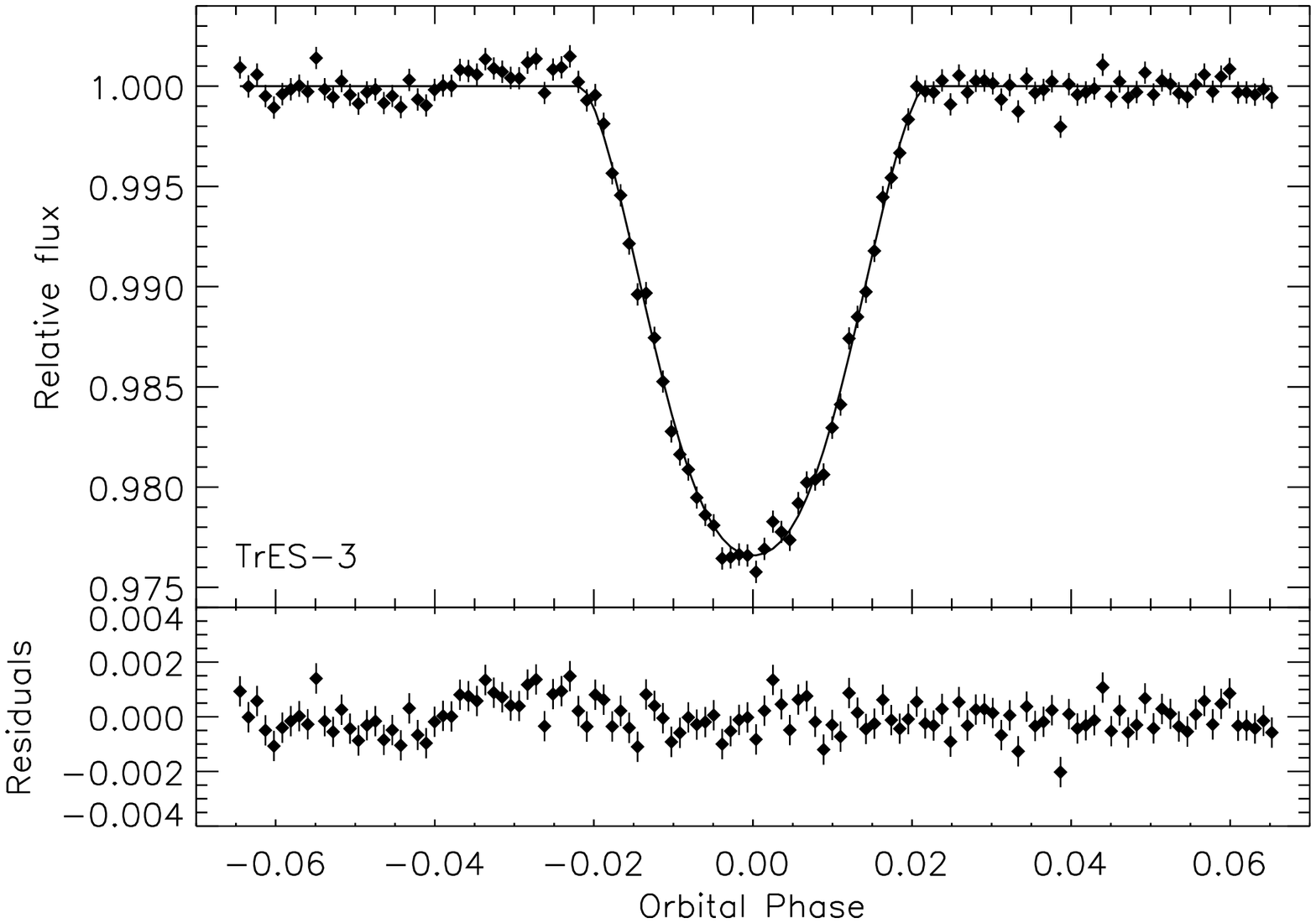}
 \caption{{\it Upper panel}: The seven TrES-3 transits, phase-folded and binned in two minute intervals. The solid line is the best-fit transit model. {\it Lower panel}: The residuals when the best-fit model is subtracted from the data.}
   \label{fig:tres3_phased}
\end{center}
\end{figure}

\begin{figure}[h!]
\begin{center}
 \includegraphics[width=4in, angle=90]{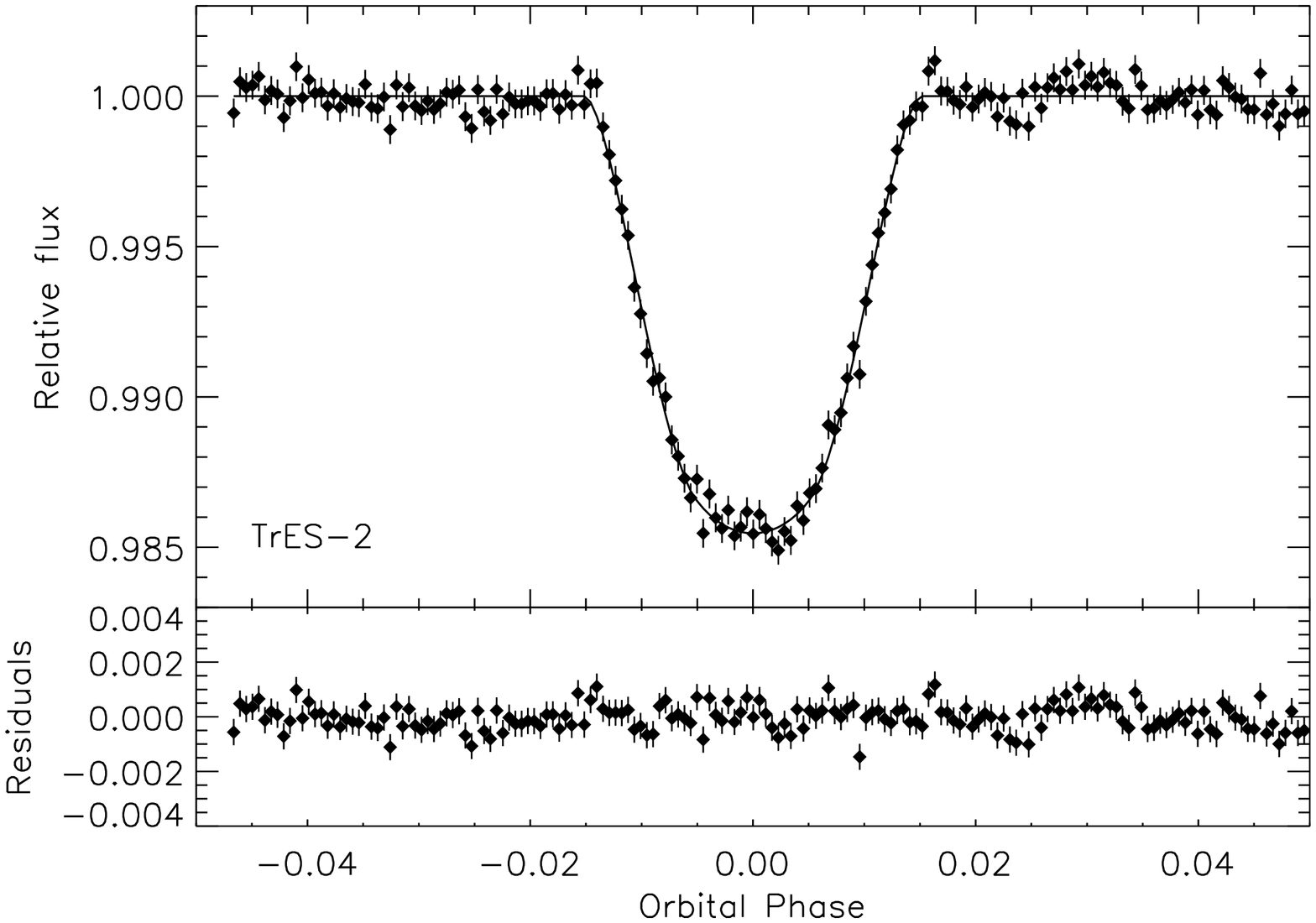}
 \caption{{\it Upper panel}: The nine TrES-2 transits, phase-folded and binned in two minute intervals. The solid line is the best-fit transit model. {\it Lower panel}: The residuals when the best-fit model is subtracted from the data.}
   \label{fig:tres2_phased}
\end{center}
\end{figure}

\begin{figure}[h!]
\begin{center}
 \includegraphics[width=4in, angle=90]{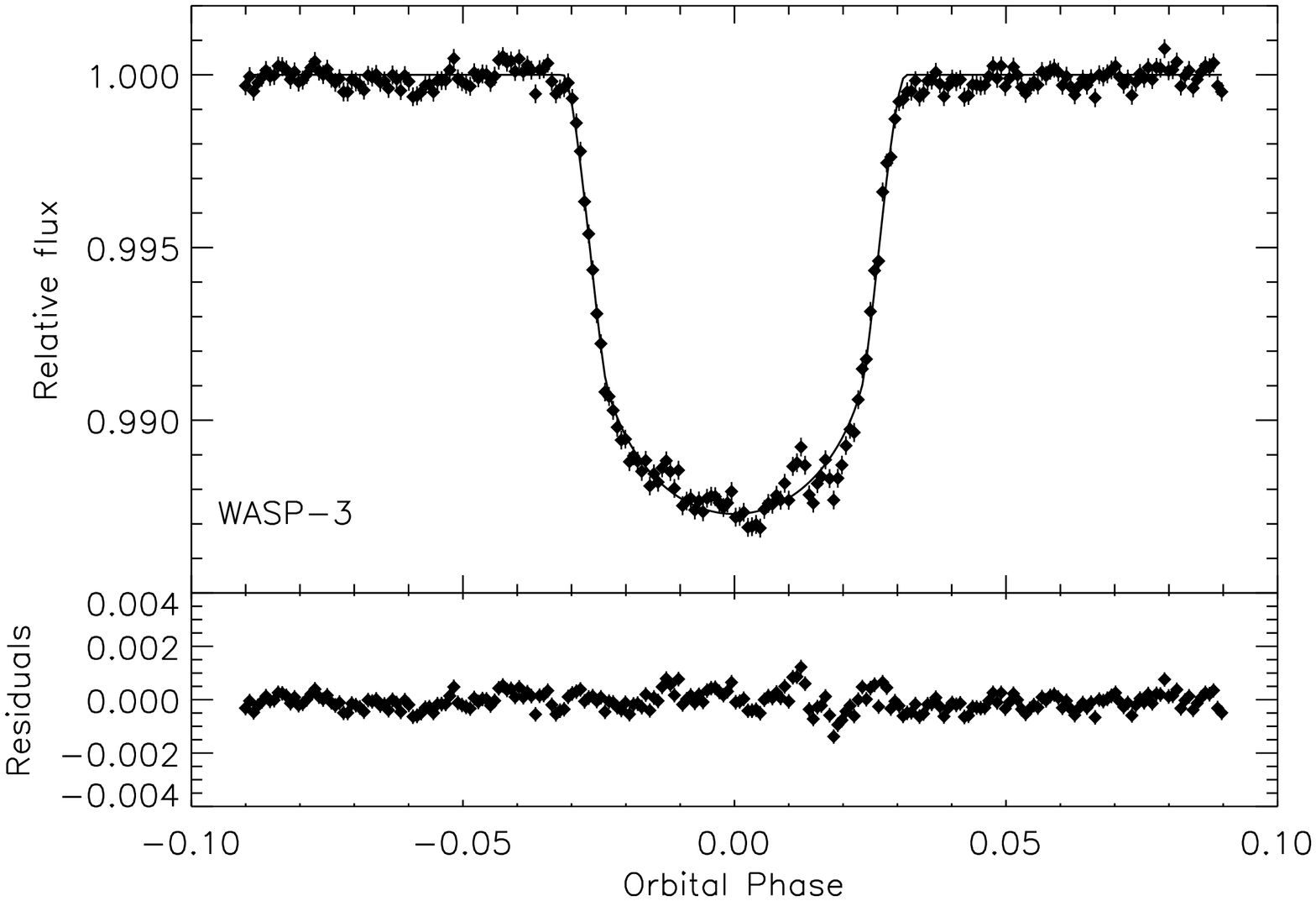}
 \caption{{\it Upper panel}: The eight WASP-3 transits, phase-folded and binned in two minute intervals. The solid line is the best-fit transit model. {\it Lower panel}: The residuals when the best-fit model is subtracted from the data. The significant in-transit deviation from the model is discussed in Section \ref{sec:disc}.}
   \label{fig:wasp3_phased}
\end{center}
\end{figure}

When fitting the transits, we use the analytic equations of \citet{Mandel02} to generate a model transit, and use $\chi^2$ as a goodness-of-fit estimator. We use the Levenberg-Marquardt algorithm to fit three dimensionless geometric parameters of the system: $R_p/R_{\star}$, $R_{\star}/a$ and cos $i$, where $R_p$ is the planetary radius, $R_{\star}$ is the stellar radius, $a$ is the semi-major axis of the planetary orbit and $i$ is the inclination of the orbit. We fix the period to the published value, but allow the time of center of transit to vary independently for each of the transits. We then use the published mass values for each of the systems to convert the transit parameters to physical properties, drawing values from \citet{Kovacs07} for HAT-P-4, \citet{Sozzetti09} for TrES-3, \citet{Sozzetti07} for TrES-2 and \citet{Pollacco08} for WASP-3. The final results of these fits are given in Table \ref{tab:hatp4} for HAT-P-4, Table \ref{tab:tres3} for TrES-3, Table \ref{tab:tres2} for TrES-2 and Table \ref{tab:wasp3} for WASP-3. We also give the transit duration from first to fourth contact for each best-fit model. For WASP-3, we discard the final transit (which was significantly offset in flux due to correlated noise), and also a partial transit (which included only the ingress), for a total of six transits. The phase-folded and binned transits for each target are shown in Figure \ref{fig:hatp4_phased} for HAT-P-4, Figure \ref{fig:tres3_phased} for TrES-3, Figure \ref{fig:tres2_phased} for TrES-2 and Figure \ref{fig:wasp3_phased} for WASP-3. 

The errors on the parameters are calculated using the residual permutation ``rosary bead'' method \citep{Winn08}. For each target, we find the residuals to the best-fit model. We shift these residuals forward collectively to the next time stamp and add the best fit models back to the new residuals, generating a new realization of the light curve which retains the correlated noise signals in the original light curve. We repeat this process 8000 times (covering approximately six days) and each time we fit for and record the geometric parameters, times of center of transit, and limb-darkening coefficients where appropriate. For each parameter we construct a histogram of the 8000 measurements, to which we fit a Gaussian. We then define the error on that parameter by the half-width half-maximum value of the best fit Gaussian. We find that increasing the number of iterations beyond 4000 does not significantly change the calculated errors.

To find the errors in the transit times, we perform a second rosary bead analysis, holding the geometric and limb-darkening values fixed and allowing only the times of center of transit to vary. We find that 4000 iterations are sufficient to sample the range of correlated noise signals, and calculate the errors in the same fashion as the geometric parameters. For each target we calculate a new orbital period and epoch by performing a weighted linear fit to the EPOCh transit times and any published transit times.

\begin{deluxetable}{lc}
\tabletypesize{\scriptsize}
\tablecaption{HAT-P-4 system parameters}
\tablewidth{0pt}
\tablehead{\colhead{Parameter} & \colhead{Value}}
\startdata
Adopted values\tablenotemark{a} & \\
$M_{\star}$ $(M_{\odot})$ & $1.26\pm0.14$ \\
$M_p$ $(M_{\rm Jup})$ & $0.68\pm0.04$ \\
\\
Transit fit values & \\
$R_p/R_{\star}$ & $0.0855\pm0.0078$\\
$a/R_{\star}$ & $0.1672\pm0.0078$\\
$i$ (deg) & $89.67\pm0.30$ \\
\\
Derived values & \\
$P$ (days) & $3.0565114\pm0.0000028$ \\
$T_c$ (BJD) & $2,454,502.56227\pm0.00021$ \\
$R_{\star}$ ($R_{\odot}$) & $1.602\pm0.061$ \\
$R_p$ ($R_{\rm Jup}$) & $1.332\pm0.052$ \\
$\tau$ (mins) & $255.6\pm1.9$ \\
\\
Limb-darkening coefficients & \\
$a$ & 0.314 \\
$b$ & 0.366 \\
\\
Transit times (BJD) &  $  2,454,490.33445\pm 0.00072$\\
&  $  2,454,493.39232\pm 0.00061$\\
&  $  2,454,496.44984\pm 0.00056$\\
&  $  2,454,499.50426\pm 0.00070$\\
&  $  2,454,502.56156\pm 0.00056$\\
&  $  2,454,505.62006\pm 0.00082$\\
&  $  2,454,508.67569\pm 0.00056$\\
&  $  2,454,649.27624\pm 0.00064$\\
&  $  2,454,652.33053\pm 0.00065$\\
&  $  2,454,655.38842\pm 0.00065$\\
\enddata
\tablenotetext{a}{Masses are from \cite{Kovacs07}.}
\label{tab:hatp4}
\end{deluxetable}

\begin{deluxetable}{lc}
\tabletypesize{\scriptsize}
\tablecaption{TrES-3 system parameters}
\tablewidth{0pt}
\tablehead{\colhead{Parameter} & \colhead{Value}}
\startdata
Adopted values\tablenotemark{a} & \\
$M_{\star}$ $(M_{\odot})$ & $0.928^{+0.028}_{-0.048}$ \\
$M_p$ $(M_{\rm Jup})$ & $1.910^{+0.075}_{-0.080}$ \\
\\
Transit fit values & \\
$R_p/R_{\star}$ & $0.1661\pm0.0343$\\
$a/R_{\star}$ & $0.1664\pm0.0204$\\
$i$ (deg) & $81.99\pm0.30$ \\
\\
Derived values & \\
$P$ (days) & $1.30618608\pm0.00000038$ \\
$T_c$ (BJD) & $2,454,538.58069\pm0.00021$ \\
$R_{\star}$ ($R_{\odot}$) & $0.817\pm0.022$ \\
$R_p$ ($R_{\rm Jup}$) & $1.320\pm0.057$ \\
$i$ (deg) & $81.99\pm0.30$ \\
$\tau$ (mins) & $81.9\pm1.1$ \\
\\
Limb-darkening coefficients & \\
$c_1$ & 0.5169\\
$c_2$ & -0.6008\\
$c_3$ & 1.4646\\
$c_4$ & -0.5743\\
\\
Transit times (BJD) &  $2,454,532.04939  \pm 0.00033$\\
&  $2,454,533.35515  \pm 0.00035$\\
&  $2,454,537.27463  \pm 0.00038$\\
&  $2,454,538.58126  \pm 0.00035$\\
&  $2,454,539.88703  \pm 0.00040$\\
&  $2,454,541.19261  \pm 0.00035$\\
&  $2,454,542.49930  \pm 0.00041$\\
\enddata 
\tablenotetext{a}{Masses are from \cite{Sozzetti09}.}
\label{tab:tres3}
\end{deluxetable}

\begin{deluxetable}{lc}
\tabletypesize{\scriptsize}
\tablecaption{TrES-2 system parameters}
\tablewidth{0pt}
\tablehead{\colhead{Parameter} & \colhead{Value}}
\startdata
Adopted values\tablenotemark{a} & \\
$M_{\star}$ $(M_{\odot})$ & $0.98\pm0.062$ \\
$M_p$ $(M_{\rm Jup})$ & $1.198\pm0.053$ \\
\\
Transit fit values & \\
$R_p/R_{\star}$ & $0.1278\pm0.0094$\\
$a/R_{\star}$ & $0.1230\pm0.0179$\\
$i$ (deg) & $84.15\pm0.16$ \\
\\
Derived values & \\
$P$ (days) & $2.47061344\pm0.0000075$ \\
$T_c$ (BJD) & $2,454,4664.23039\pm0.00018$ \\
$R_{\star}$ ($R_{\odot}$) & $0.940\pm0.026$ \\
$R_p$ ($R_{\rm Jup}$) & $1.169\pm0.034$ \\
$\tau$ (mins) & $107.3\pm1.1$ \\
\\
Limb-darkening coefficients & \\
$c_1$ & 0.3899\\
$c_2$ & -0.1391\\
$c_3$ & 0.9662\\
$c_4$ & -0.0329\\
\\
Transit times (BJD) &  $2,454,646.93735  \pm 0.00032$\\
&  $2,454,656.81879  \pm 0.00034$\\
&  $2,454,659.28871  \pm 0.00042$\\
&  $2,454,661.76005  \pm 0.00044$\\
&  $2,454,664.23072  \pm 0.00050$\\
&  $2,454,669.17156  \pm 0.00028$\\
&  $2,454,671.64117  \pm 0.00028$\\
&  $2,454,674.11318  \pm 0.00033$\\
&  $2,454,676.58257  \pm 0.00051$\\
\enddata
\tablenotetext{a}{Masses are from \cite{Sozzetti07}.}
\label{tab:tres2}
\end{deluxetable}

\begin{deluxetable}{lc}
\tabletypesize{\scriptsize}
\tablecaption{WASP-3 system parameters}
\tablewidth{0pt}
\tablehead{\colhead{Parameter} & \colhead{Value}}
\startdata
Adopted values\tablenotemark{a} & \\
$M_{\star}$ $(M_{\odot})$ & $1.24^{+0.06}_{-0.11}$ \\
$M_p$ $(M_{\rm Jup})$ & $1.76^{+0.08}_{-0.14}$ \\
\\
Transit fit values & \\
$R_p/R_{\star}$ & $0.1051\pm0.0124$\\
$a/R_{\star}$ & $0.1989\pm0.0287$\\
$i$ (deg) & $84.15\pm0.16$ \\
\\
Derived values & \\
$P$ (days) & $1.8468373\pm0.0000014$ \\
$T_c$ (BJD) & $2,454,686.82069\pm0.00039$ \\
$R_{\star}$ ($R_{\odot}$) & $1.354\pm0.056$ \\
$R_p$ ($R_{\rm Jup}$) & $1.385\pm0.060$ \\
$i$ (deg) & $84.22\pm0.81$ \\
$\tau$ (mins) & $167.3\pm1.3$ \\
\\
Limb-darkening coefficients & \\
$c_1$ & 0.2185\\
$c_2$ & 0.6183\\
$c_3$ & -0.1040\\
$c_4$ & -0.0426\\
\\
Transit times (BJD) &  $2,454,679.43264  \pm 0.00050$\\
&  $2,454,681.27911  \pm 0.00040$\\
&  $2,454,683.12740  \pm 0.00035$\\
&  $2,454,684.97486  \pm 0.00027$\\
&  $2,454,686.82053  \pm 0.00059$\\
&  $2,454,690.51381  \pm 0.00055$\\
&  $2,454,692.36117  \pm 0.00043$\\
&  $2,454,694.20711  \pm 0.00042$\\
\enddata
\tablenotetext{a}{Masses are from \cite{Pollacco08}.}
\label{tab:wasp3}
\end{deluxetable}

\section{Secondary eclipse constraints}
\label{sec:eclipses}

Our constraints of the secondary eclipse depths are limited by the correlated noise in the {\it EPOXI} data. Ideally, for each target we would combine our multiple observations of the secondary eclipses to increase the signal to noise. However, the fluctuations due to the correlated noise preclude this. For example, Figure \ref{fig:tres3_eclipses} shows six of the TrES-3 secondary eclipses, where in some cases correlated noise results in an increase in flux at the time of secondary eclipse, instead of the expected decrement. If we assume that the secondary eclipse in the EPOCh bandpass, with a central wavelength of 650~nm, is due exclusively to the reflected light of the planet, then the eclipse depths we would anticipate, for a geometric albedo of 1, would range from 0.02\% for HAT-P-4 to 0.08\% for TrES-3.\footnote{In fact, the CCD is found to be quite efficient at the redder wavelengths, and it is therefore feasible that for the hottest planets there may be a contribution from the thermal emission of the planet, resulting in deeper secondary eclipses.} 

Since the fluctuations from correlated noise in the measured eclipse depths are sometimes larger than the signal we expect to measure, we choose not to combine the multiple observations and instead analyze each eclipse independently. Our intent is to use the scatter of individual eclipse measurements to constrain the amplitude of the correlated noise. As for the transits, for each eclipse in the data we apply a point-by-point correction to the data in and adjacent to the eclipse. For the targets presented here we assume that $e=0$ and therefore that the secondary eclipse occurs at a phase of 0.5. For TrES-3 and TrES-2 this assumption is strongly supported by previous secondary eclipse measurements with the {\it Spitzer} IRAC instrument which demonstrated no evidence of non-zero eccentricity (Fressin et al. 2010 and O'Donovan et al. 2010 respectively). For HAT-P-4 and WASP-3, the extant radial velocity data are consistent with circular orbits (Kovacs et al. 2007 and Pollacco et al. 2008 respectively). In addition to the point-by-point correction, we fit a linear time-dependent slope to the adjacent out-of-eclipse data to remove any remaining long timescale trends. Finally, we separate the data into 10 minute bins and remove 3$\sigma$ flux outliers from each bin. For TrES-2, we discard the first observed eclipse, which was obtained during the pre-look for this target, since the pre-look data are not well calibrated by the surface spline generated for the remaining data. We assume this is due to changes in the CCD in the time that occurred between the pre-look and the full set of observations. We also discard eclipses where less than half the event is observed, one for TrES-2, one for WASP-3 and one for HAT-P-4. As discussed in Section \ref{sec:details}, we finally discard four of the nine WASP-3 secondary eclipses that fall on regions of the CCD we cannot calibrate.

We fit the eclipses using a transit model with the best-fit parameters from the transit analysis and no limb-darkening. We then scale the depth of this model to fit the data, finding the depth that minimizes the $\chi^2$ value. For each target we then find the mean ($\bar{x}$) and standard deviation ($\sigma_x$) of the individual best-fit depths, and define the 95\% confidence upper limit on the eclipse depth as $\bar{x}+2\sigma_x$. The secondary eclipses of HAT-P-4, TrES-3, TrES-2 and WASP-3 are shown in Figures 9--12. The upper limits are given in Table \ref{tab:eclipses}. We note that we achieve a useful constraint only in the case of TrES-3.


\begin{deluxetable}{lccc}
\tabletypesize{\scriptsize}
\tablecaption{EPOCh secondary eclipse measurements}
\tablewidth{0pt}
\tablehead{\colhead{Target} & \colhead{Eclipse Depth} & \colhead{Upper limit} & \colhead{Implied $A_g$}}
\startdata
HAT-P-4 & $-0.0069\pm0.0397$\% & 0.073\% & 3.5 \\
TrES-3  & $-0.020\pm0.041$\%   & 0.062\% & 0.81 \\
TrES-2  & $0.023\pm0.071$\%    & 0.16\% & 6.6 \\
WASP-3  & $0.023\pm0.044$\%    & 0.11\% & 2.5 \\
\enddata
\label{tab:eclipses}
\end{deluxetable}

\begin{figure}[h!]
\begin{center}
 \includegraphics[width=4in]{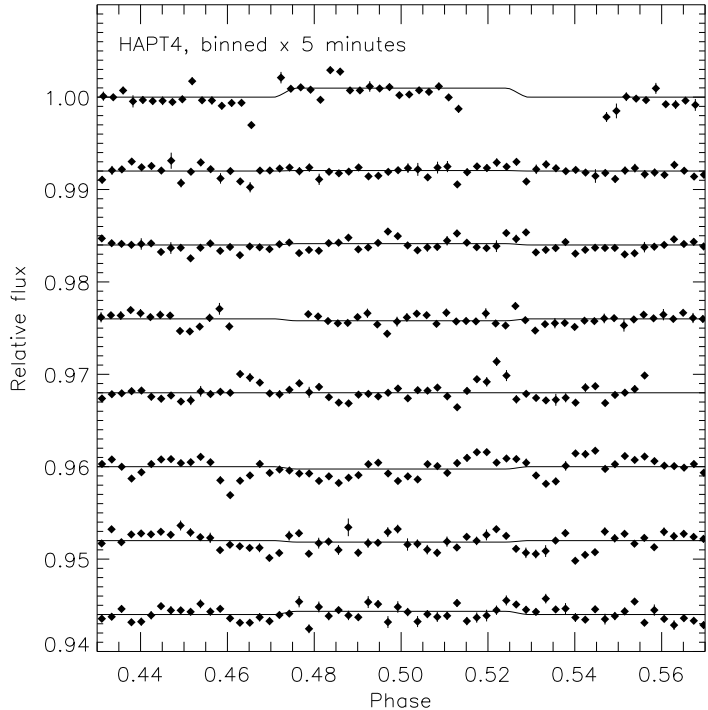}
 \caption{Eight EPOCh secondary eclipse observations of HAT-P-4, offset in relative flux for clarity and binned in five minute intervals. The error on each point is $\sigma/\sqrt{N}$, where $\sigma$ is the scatter in the bin and $N$ the number of points. The solid lines are the best fit eclipse model in each case. The bottom three eclipses were obtained in the contingent block of observations.}
   \label{fig:hatp4_eclipses}
\end{center}
\end{figure}

\begin{figure}[h!]
\begin{center}
 \includegraphics[width=4in]{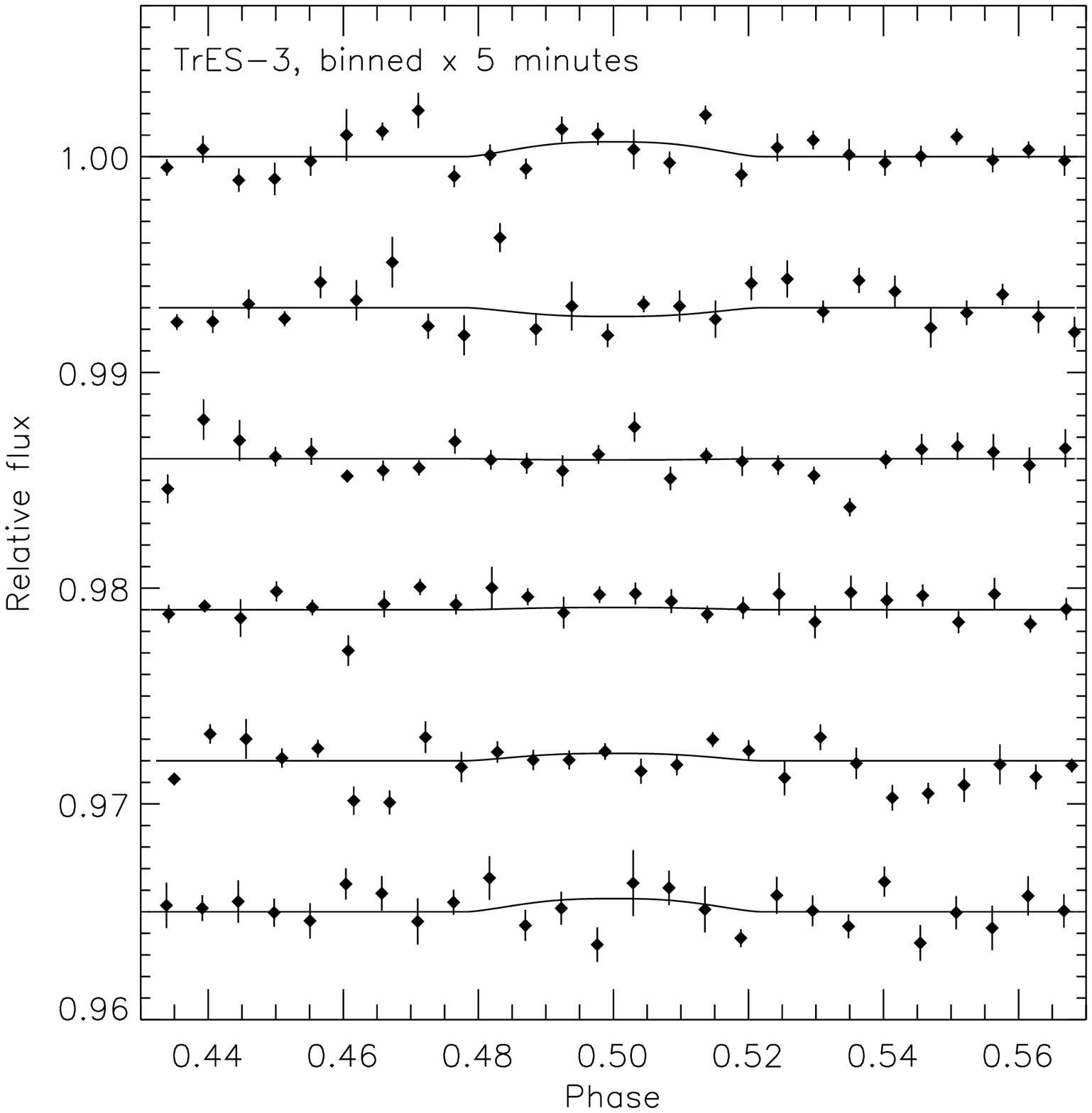}
 \caption{Six EPOCh secondary eclipse observations of TrES-3, offset in relative flux for clarity. The error on each point is $\sigma/\sqrt{N}$, where $\sigma$ is the scatter in the bin and $N$ the number of points. The solid lines are the best fit eclipse model in each case.}
   \label{fig:tres3_eclipses}
\end{center}
\end{figure}

\begin{figure}[h!]
\begin{center}
 \includegraphics[width=4in, angle=90]{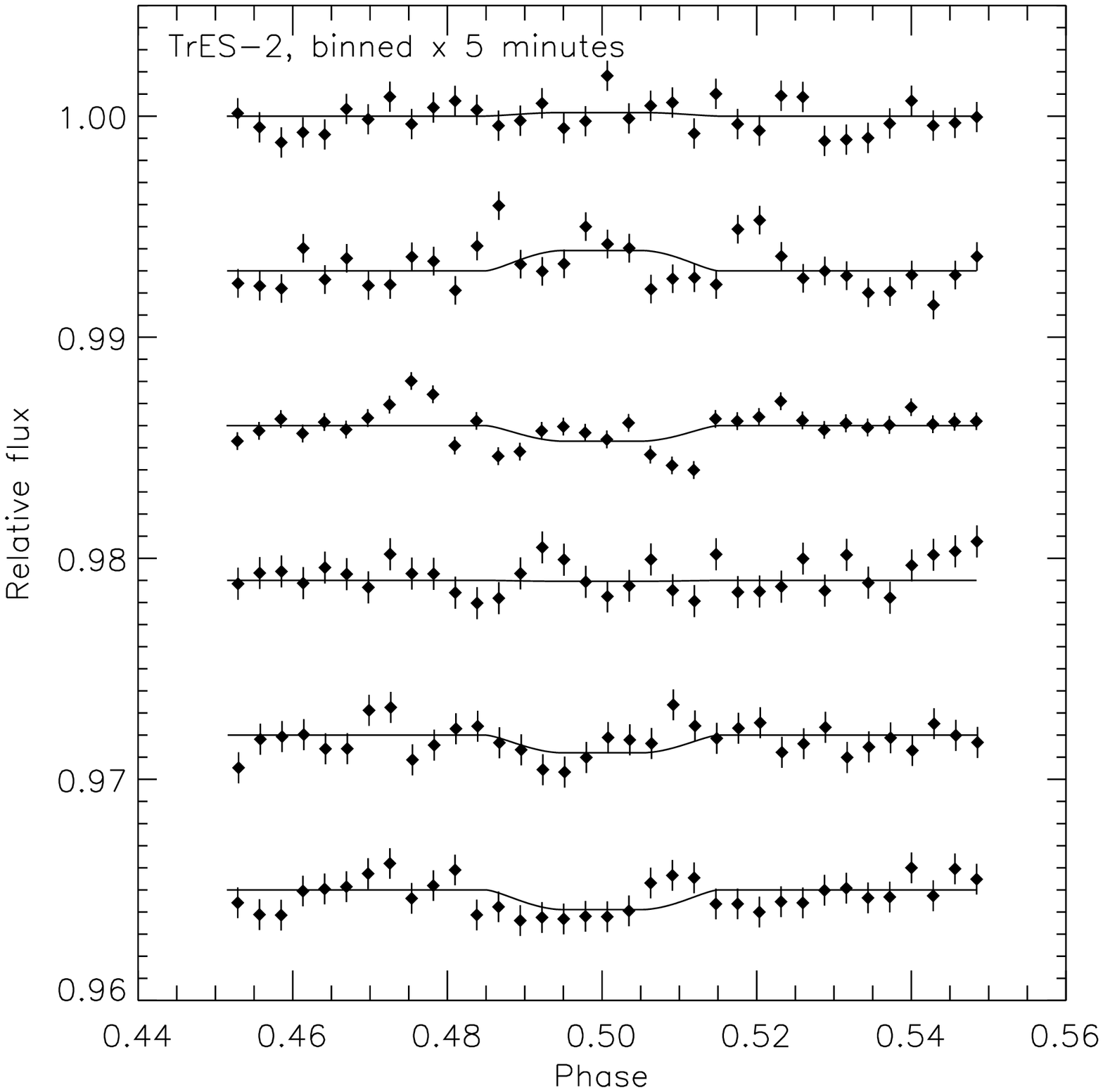}
 \caption{Six EPOCh secondary eclipse observations of TrES-2, offset in relative flux for clarity. The error on each point is $\sigma/\sqrt{N}$, where $\sigma$ is the scatter in the bin and $N$ the number of points. The solid lines are the best fit eclipse model in each case.}
   \label{fig:tres2_eclipses}
\end{center}
\end{figure}

\begin{figure}[h!]
\begin{center}
 \includegraphics[width=4in, angle=90]{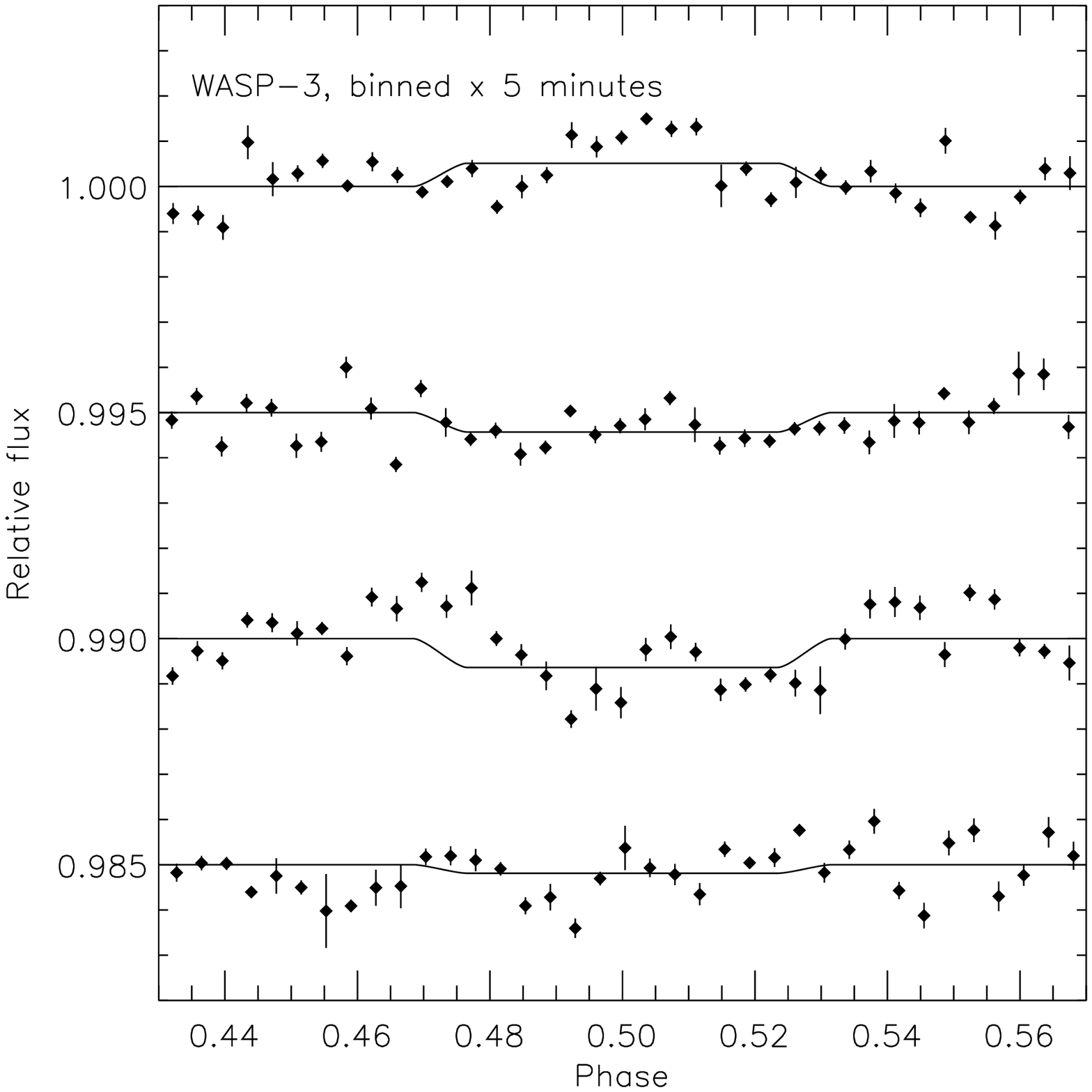}
 \caption{Four EPOCh secondary eclipse observations of WASP-3, offset in relative flux for clarity. The error on each point is $\sigma/\sqrt{N}$, where $\sigma$ is the scatter in the bin and $N$ the number of points. The solid lines are the best fit eclipse model in each case.}
   \label{fig:wasp3_eclipses}
\end{center}
\end{figure}

\section{Discussion}
\label{sec:disc}

\subsection{HAT-P-4}
\label{sec:hatp4}

\begin{figure}[h!]
\begin{center}
 \includegraphics[width=4in, angle=90]{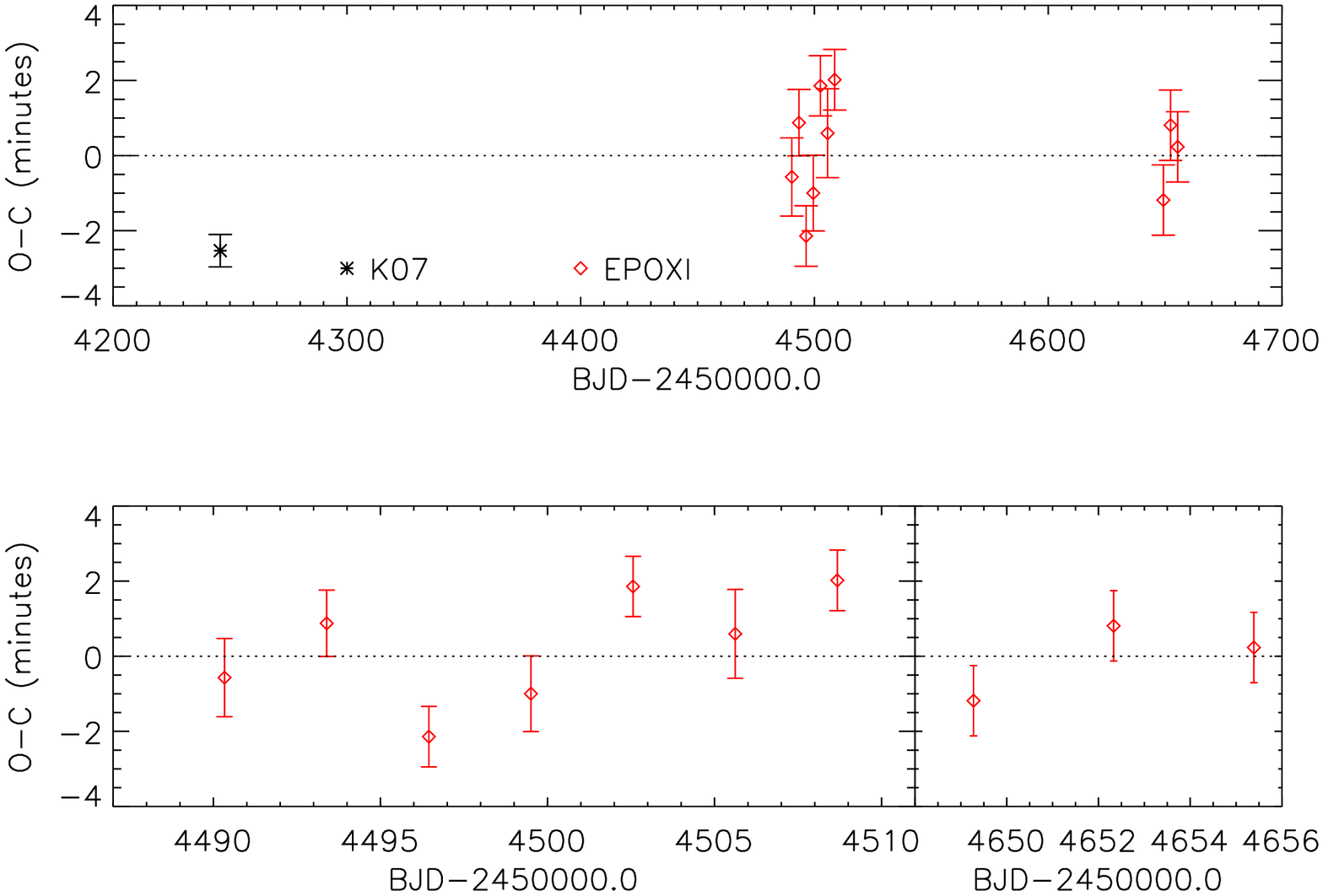}
 \caption{The transit times of HAT-P-4. The open diamonds are the EPOCh transit times from this paper; the asterisk is the discovery epoch \citep{Kovacs07}. {\it Lower panel}: An expanded view of the EPOCh transit times, with $1\sigma$ errors of 48-71 seconds.}
   \label{fig:hatp4_ttvs}
\end{center}
\end{figure}

For HAT-P-4, our estimates of the system parameters are consistent with (and in the case of inclination, more precise than) those published by \citet{Kovacs07} and \citet{Torres08}. We calculate $R_p=1.332\pm0.052$ $R_{\rm Jup}$, $R_{\star}=1.602\pm0.061$ $R_{\odot}$, $i=89.67\pm0.30$ degrees and $\tau=255.6\pm1.9$ minutes, where $\tau$ is the transit duration from first to fourth contact. We use the discovery epoch and the ten EPOCh transit times presented in this paper to produce a new refined ephemeris of $T_c({\rm BJD}) = 2454245.81531\pm0.00021 + 3.0565114\pm0.0000028E$. Figure \ref{fig:hatp4_ttvs} shows the residuals to the new ephemeris. We see no evidence for transit timing variations in the residuals which have a scatter of roughly 2 minutes.

We use eight of the nine observed secondary eclipses to constrain the depth of the eclipse in the EPOCh bandpass, discarding the ninth due to poor coverage of the event. The eclipses are shown in Figure \ref{fig:hatp4_eclipses}. We set a 95\% confidence upper limit on the eclipse depth of 0.073\%, which, if it were produced entirely by reflected light, would correspond to a planetary geometric albedo of $A_g=3.5$, a physically impossible value. In the future, full phase curves of HAT-P-4 are scheduled to be observed in the near-infrared 3.6 and 4.5 micron IRAC bands, as part of the Warm {\it Spitzer} census of exoplanet atmospheres, at which point we may begin to study the atmosphere in more detail.

\subsection{TrES-3}
\label{sec:tres3}

\begin{figure}[h!]
\begin{center}
 \includegraphics[width=4in, angle=90]{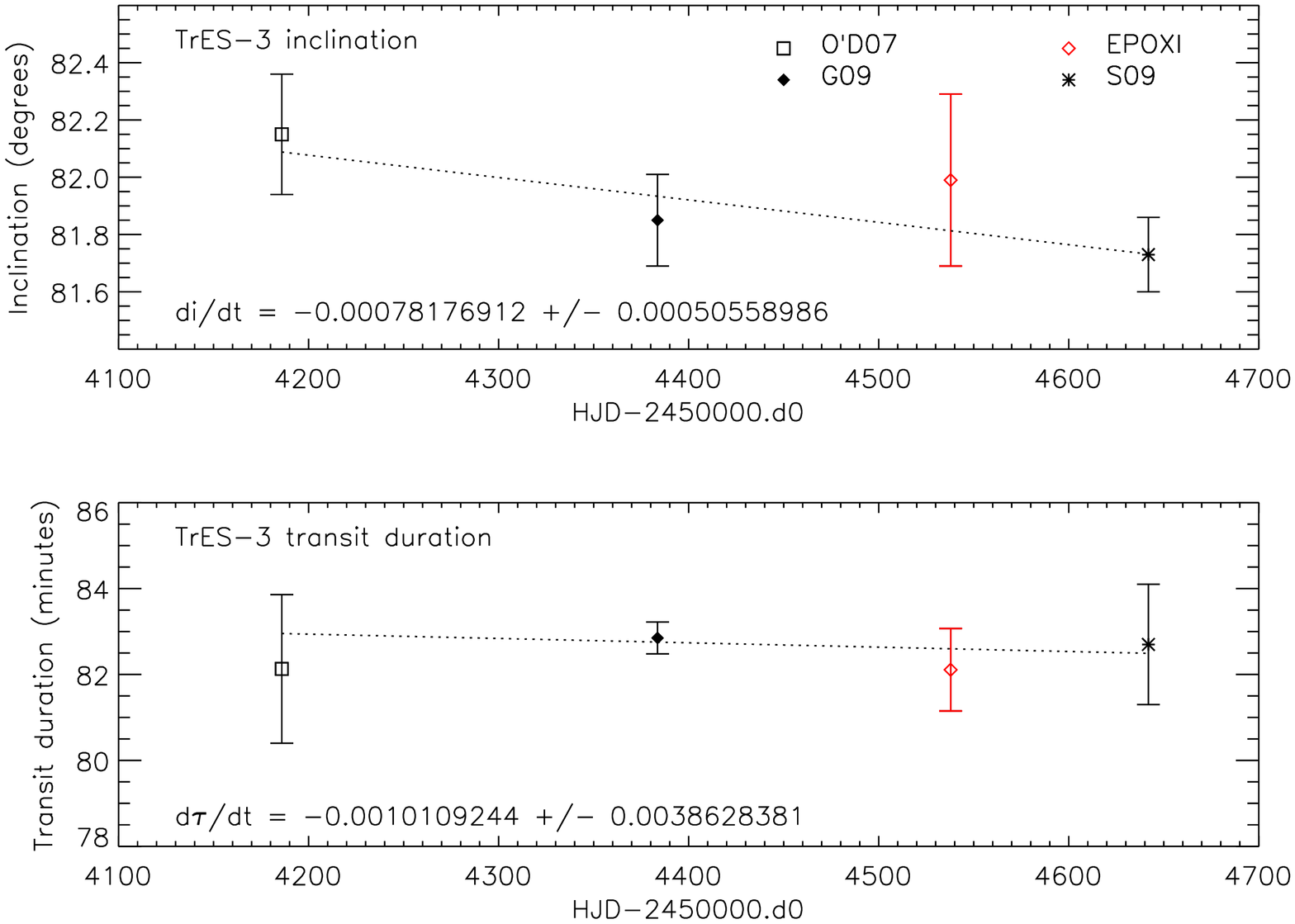}
 \caption{{\it Upper panel}: The estimates of the inclination for TrES-3 as a function of time. {\it Lower panel}: The estimates of the TrES-3 transit durations.}
   \label{fig:tres3_trends}
\end{center}
\end{figure}

For TrES-3, we find system parameters consistent with those published by \citet{ODonovan07}, \citet{Sozzetti09} and \citet{Gibson09}, with $R_p=1.320\pm0.057$ $R_{\rm Jup}$, $R_{\star}=0.817\pm0.022$ $R_{\odot}$, $i=81.99\pm0.30$ degrees and $\tau=81.9\pm1.1$ minutes. In the upper panel of Figure \ref{fig:tres3_trends} we plot the published values of inclination with time and note a weak trend towards decreasing inclination, however it is present at only the 1.5$\sigma$ level, and hence not significant (and largely dependent on the most recent value from Sozzetti et al. 2009). A more model-independent way of constraining changes in the transit parameters with time is by measuring the transit duration. Where available, we use the quoted transit duration and error, and otherwise we calculate the transit duration from the published parameters, using equation (4) from \citet{Charbonneau06}. Following the analytic approximation of \citet{Carter08}, we set the error on these calculated transit durations to twice the error in the measured transit times for each source. Although this error was originally derived for the transit duration from mid-ingress to mid-egress, as compared to the transit duration from first to fourth contact, we find that for the {\it EPOCh} data the errors calculated using this approximation and the errors measured from the data themselves are nearly identical (1.1 minutes and 1.0 minutes respectively). We plot the derived values in the lower panel of Figure \ref{fig:tres3_trends}, and we see no evidence of a change in the transit duration with time.

In Section \ref{sec:details} we noted that in the process of calibrating the light curve, a long term variability was evident. This variability is consistent with stellar variability due to spots. Using a $v$sin$i$ of $<$2 km s$^{-1}$ \citep{ODonovan07}, the rotational period of TrES-3 must be $>$21 days, considerably longer than our observation span of 12 days. We can therefore not place any additional constraints on the rotational period of TrES-3, however we note that if the variability is due to spots on the stellar surface rotating in and out of view then additional monitoring of TrES-3 may reveal the rotational period. 

\begin{figure}[h!]
\begin{center}
 \includegraphics[width=4in, angle=90]{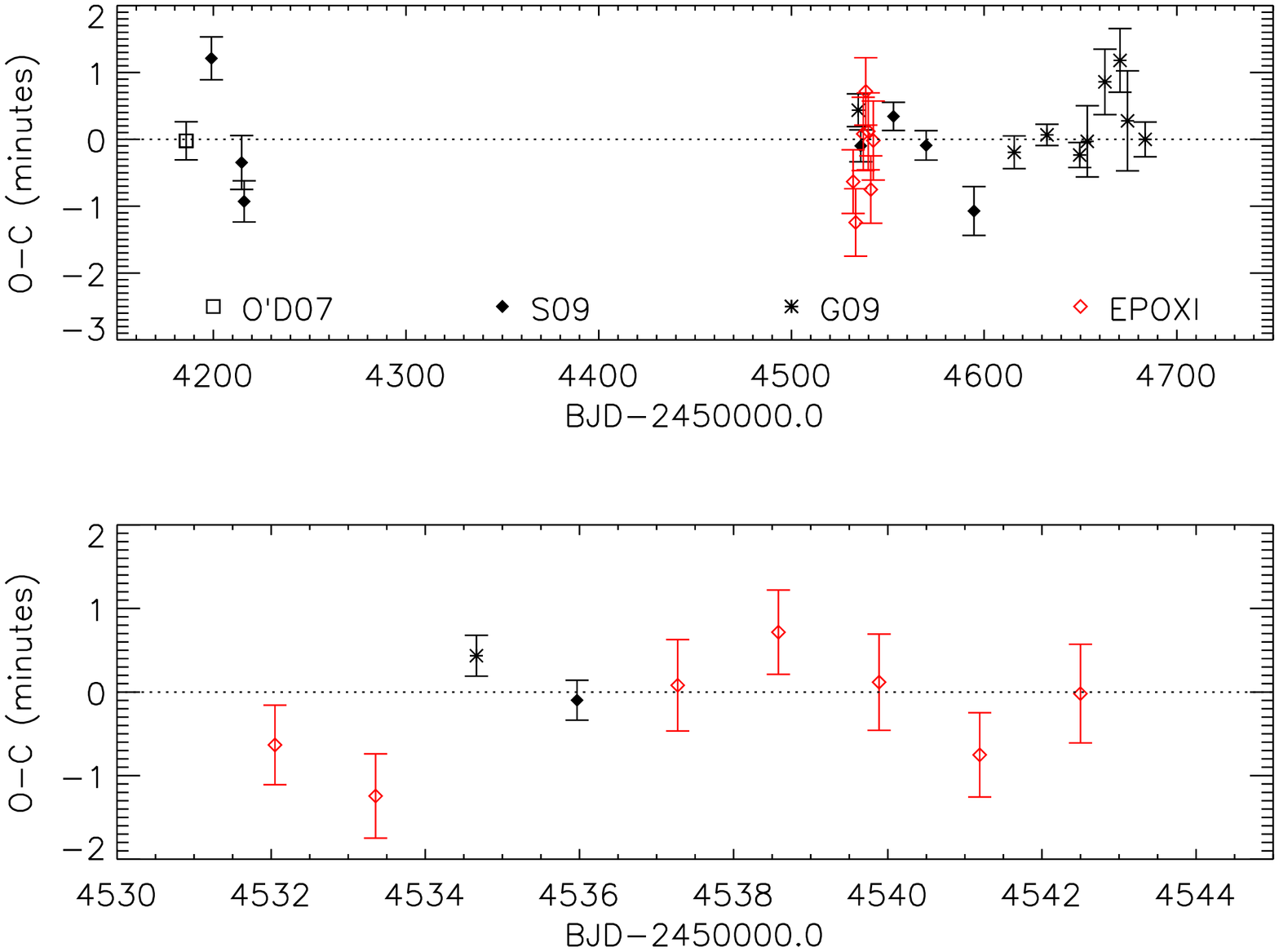}
 \caption{The times of transit of TrES-3. O'D07: \citet{ODonovan07}; S09: \citet{Sozzetti09}; G09: \citet{Gibson09}; EPOXI: this paper. {\it Lower panel}: An expanded view of the EPOCh transit times, with $1\sigma$ errors of 28--35 seconds.}
   \label{fig:tres3_ttvs}
\end{center}
\end{figure}

For TrES-3, we calculate a new ephemeris of $T_c({\rm BJD})=2454538.58069\pm0.00021+1.30618606\pm0.00000038E$ using the published transit times and the seven EPOCh transits presented in this paper. Figure \ref{fig:tres3_ttvs} shows the residuals to the new ephemeris. We see no evidence of the period changing with time or transit timing variations larger than 1 minute.

\begin{figure}[h!]
\begin{center}
 \includegraphics[width=6in]{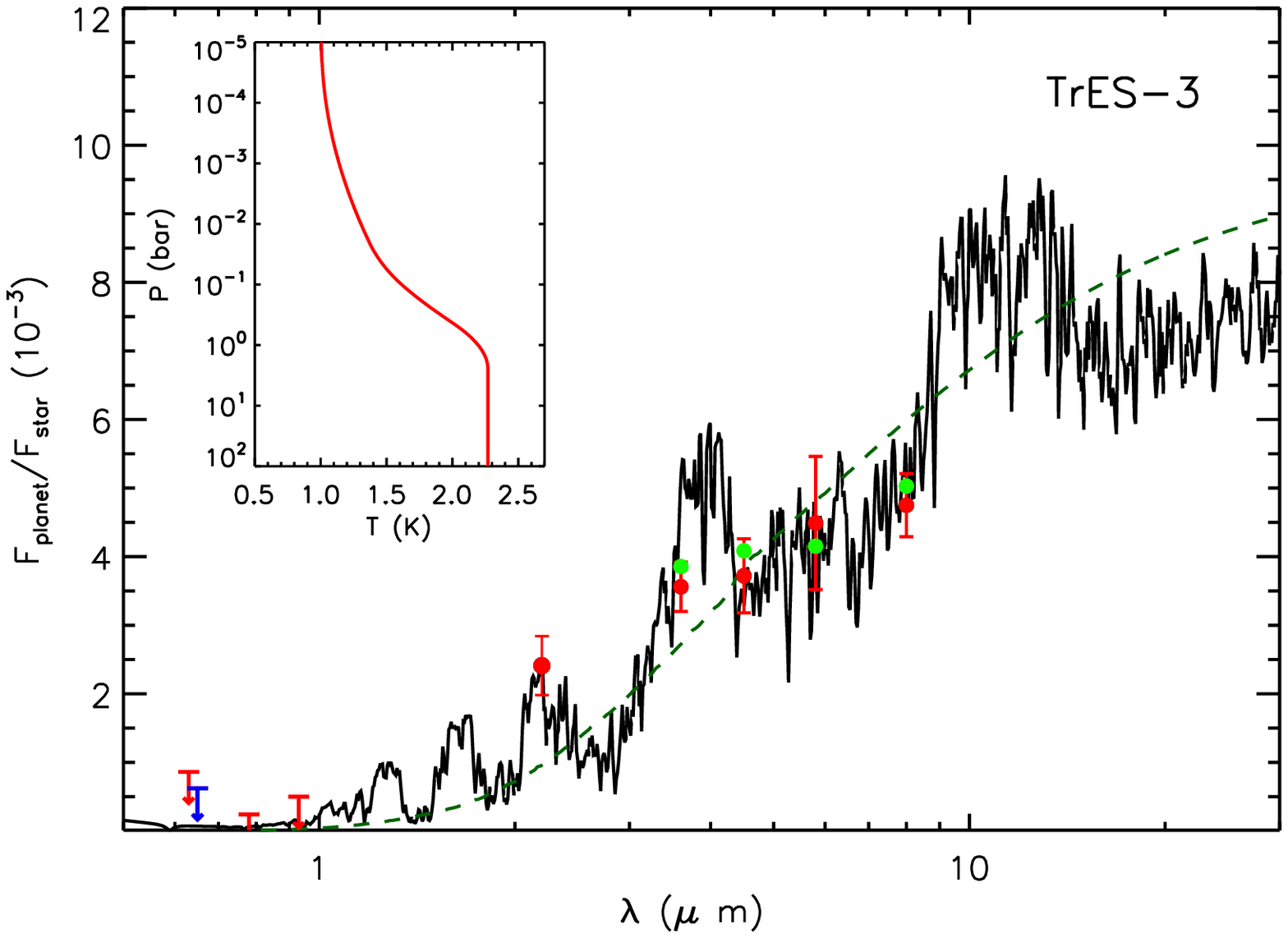}
 \caption{The optical and near-infrared secondary eclipse measurements of TrES-3. The EPOCh upper limit of 0.062\% is shown in blue at 0.65 microns. The remaining upper limits in the optical are from \citet{Winn08}; the measurement at 2.2 microns is from \citet{deMooij09}; and the four measurements from 3.6 to 8.0 microns are from \citet{Fressin10}. The solid black line is a representative model from the set of models that fit the data to within 1$\sigma$, and the dashed lined shows a black-body spectrum for a temperature of 1600K. The green circles represent the model integrated to the {\it Spitzer} bandpasses. The inset is the temperature-pressure profile for the model shown.}
   \label{fig:tres3_spectrum}
\end{center}
\end{figure}

Using the six EPOCh secondary eclipse observations of TrES-3, shown in Figure \ref{fig:tres3_eclipses}, we set a 95\% confidence upper limit on the eclipse depth of 0.062\%. This indicates the planetary geometric albedo must be $A_g<0.81$ in the EPOCh bandpass. \citet{Winn08} observed secondary eclipses of TrES-3 in the $i$, $z$ and $R$ bands, and were able to put 99\% confidence upper limits on the eclipse depths of 0.024\%, 0.050\% and 0.086\% respectively. The EPOCh upper limit at 0.65 $\micron$ is consistent with the $R$ band upper limit. \citet{deMooij09} observed the secondary eclipse in the $K$ band and found a depth of 0.241$\pm$0.043\%. \citet{Fressin10} observed secondary eclipses of TrES-3 with the {\it Spitzer} IRAC instrument, measuring depths of 0.356$\pm$0.036\%, 0.372$\pm$0.054\%, 0.449$\pm$0.097\% and 0.475$\pm$0.046\% in the 3.6, 4.5, 5.8 and 8.0 micron bands respectively. The secondary eclipse measurements are shown in Figure \ref{fig:tres3_spectrum} as a function of wavelength. 

Given the high levels of stellar irradiation, the atmosphere of TrES-3 was anticipated to host a thermal inversion \citep{Fortney08,deMooij09}. Using all data sets, however, \citet{Fressin10} found the observations to be best fit with a dayside atmosphere model without a thermal inversion.

Our model spectra are computed using the exoplanet atmosphere model developed in \citet{Madhusudhan09}. The model consists of a line-by-line radiative transfer model, with constraints of hydrostatic equilibrium and global energy balance, and coupled to a parametric pressure-temperature (P-T) structure and parametric molecular abundances (parametrized as deviations from thermochemical equilibrium and solar abundances). Our modeling approach allows one to compute large ensembles of models, and efficiently explore the parameter space of molecular compositions and temperature structure. 

We confirm previous findings that existing detections of day-side observations can be explained to within the 1$\sigma$ uncertainties by models without thermal inversions. The black curve in Figure \ref{fig:tres3_spectrum} shows one such model spectrum, which has a chemical composition at thermochemical equilibrium and solar abundances for the elements. The model is also consistent with the EPOCh upper-limit at 0.65 microns, and with the upper-limits from \citet{Winn08}. The dark green dashed curve shows a 1600K blackbody spectrum of the planet, indicating that the data cannot be explained by a pure blackbody. The model reported here has a day-night energy redistribution fraction of 0.4, indicating very efficient redistribution. Therefore, based on previous studies and our current finding, existing data do not require the presence of a thermal inversion in TrES-3. However, a detailed exploration of the model parameter space would be needed to rule out thermal inversions with a given statistical significance \citep{Madhusudhan10}.

\subsection{TrES-2}
\label{sec:tres2}

For TrES-2, we derive system parameters that are consistent at the 1.5$\sigma$ level with estimates published by \citet{ODonovan06}, \cite{Sozzetti07}, and \citet{Holman07}, finding $R_p=1.169\pm0.034$ $R_{\rm Jup}$, $R_{\star}=0.940\pm0.026$ $R_{\odot}$, $i=84.15\pm0.16$ degrees and $\tau=107.3\pm1.1$ minutes.

\begin{figure}[h!]
\begin{center}
 \includegraphics[width=4in, angle=90]{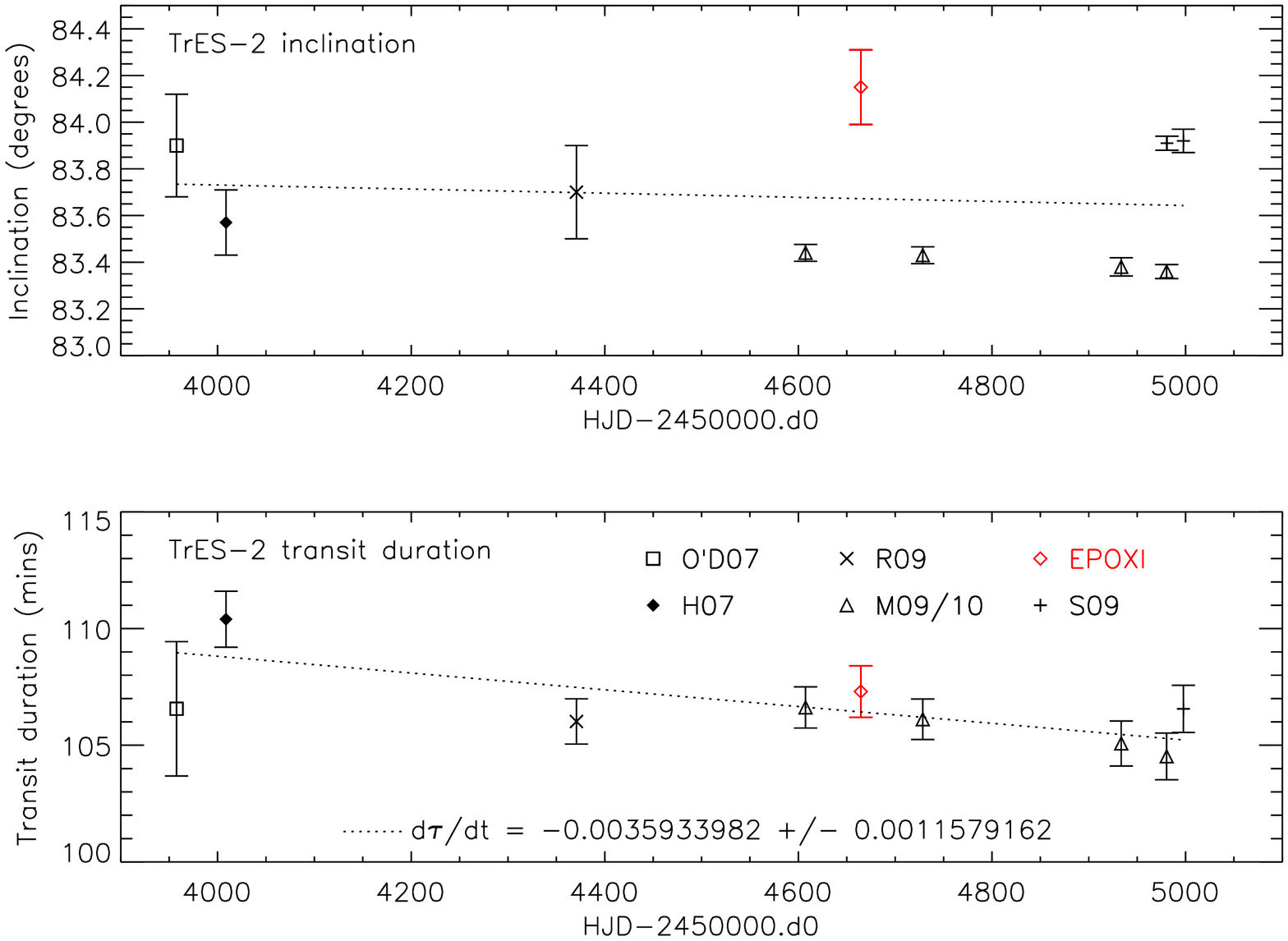}
 \caption{{\it Upper panel}: The estimates of the inclination of TrES-2 as a function of time. The dotted line is an unweighted linear fit. O'D07: \citep{ODonovan07}; H07: \citep{Holman07}; R09: \citep{Rabus09}; M09/10: \citep{Mislis09,Mislis10}; S09: \citep{Scuderi09}; EPOXI: this paper. {\it Lower panel}: The TrES-2 transit durations. In this case the dotted line is a weighted linear fit.}
   \label{fig:tres2_trends}
\end{center}
\end{figure}

As discussed in Section \ref{sec:intro}, there is currently a debate as to whether the inclination of the planetary orbit and duration of the TrES-2 transit are decreasing with time due to orbital precession. In the upper panel of Figure \ref{fig:tres2_trends} we plot the estimates  for the inclination as a function of time. For the inclination, the error bars of \citet{Mislis09}, \citet{Mislis10} and \citet{Scuderi09} were calculated by fixing the stellar and planetary radii and allowing only the inclination and time of center of transit to vary. The remainder of the inclination error bars were calculated allowing all of the geometric parameters to vary simultaneously, which explains why they are considerably larger than the later results. Since the errors skew any weighted linear fit towards an unrealistically large {\it increase} in the inclination with time, we instead plot an unweighted linear fit to guide the eye. We note that TrES-2 is in the {\it Kepler} field and that any change in inclination with time will soon be measured with exquisite precision.

The inclination measured from a particular transit light curve will necessarily depend on the geometric parameters and to some extent the choice of limb-darkening treatment. However, the transit duration is directly measurable from the light curve and should not depend on the limb darkening. The lower panel of Figure \ref{fig:tres2_trends} shows the published transit durations as a function of time. Where they were not given, we calculated the durations and errors as described for TrES-3. In this case, we perform a weighted linear fit and do see a formally significant decrease in the transit duration with time. However, this conclusion is heavily dependent on one point, in this case the duration calculated from \citet{Holman07}. If this point is excluded from the fit, then $d\tau/dt=-0.0015\pm0.0015$, consistent with no change in the transit duration with time and therefore we do not claim to have detected a change in the transit duration with time. Again, we expect {\it Kepler} to provide a clear answer to this question.

\begin{figure}[h!]
\begin{center}
 \includegraphics[width=4in, angle=90]{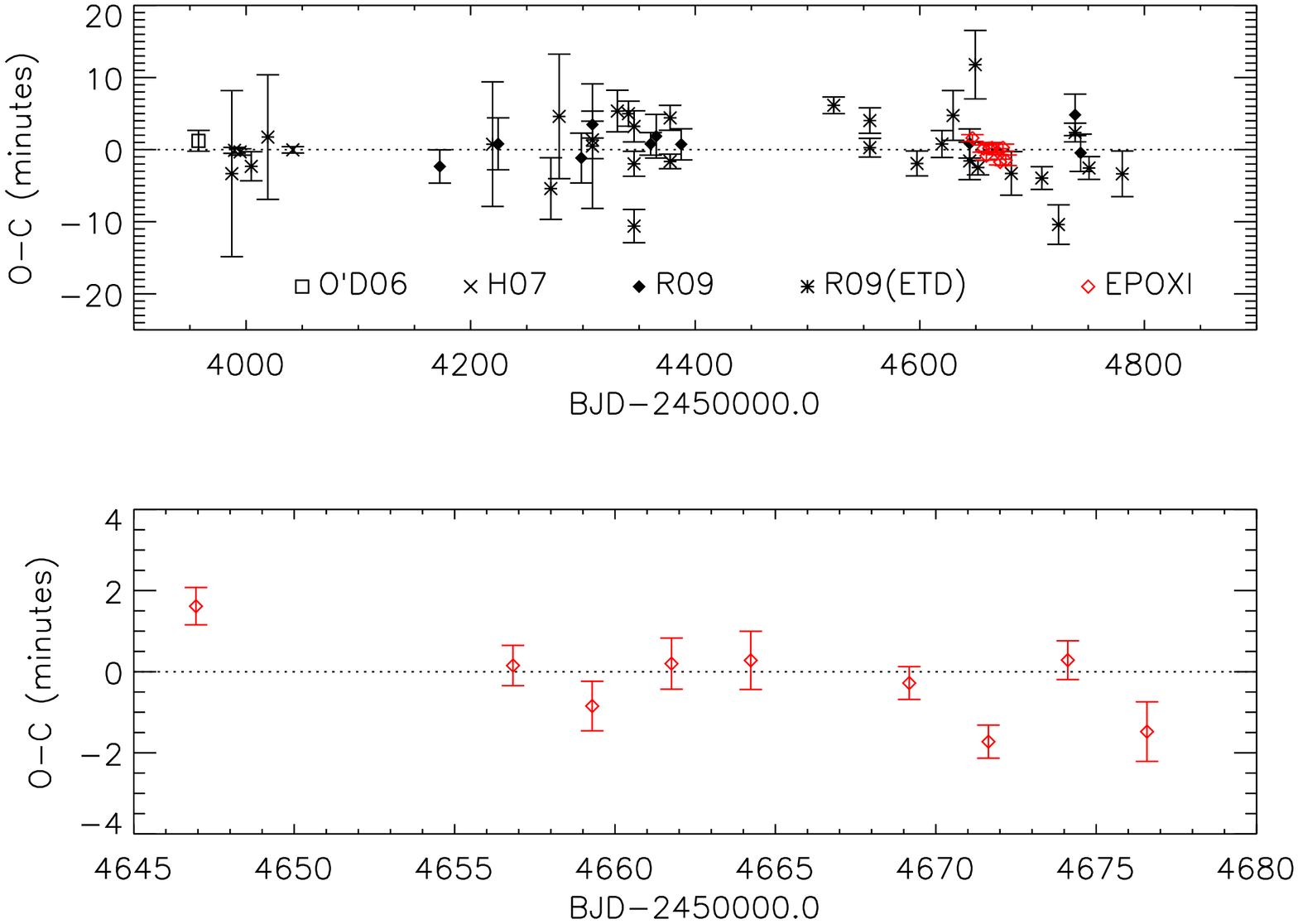}
 \caption{{\it Upper panel}: The transit times of TrES-2. O'D06: \citet{ODonovan06}; H07: \citet{Holman07}; R09: \citet{Raetz09}; R09(ETD): \citet{Raetz09} (from the Exoplanet Transit Database, http://var.astro.cz/ETD); EPOXI: this paper. {\it Lower panel}: An expanded view of the EPOCh transit times, with $1\sigma$ errors of 24--44 seconds.}
   \label{fig:tres2_ttvs}
\end{center}
\end{figure}

Using the published transit times of TrES-2 and the nine transits observed by EPOCh presented in this paper, we find a new weighted ephemeris of $T_c({\rm BJD})=24544664.23039\pm0.00018+2.47061344\pm0.00000075E$. The residuals to this ephemeris are shown in Figure \ref{fig:tres2_ttvs}. In the EPOCh residuals, we see no variations in the transit times above the level of 2 minutes; excluding the amateur data from the Exoplanet Transit Database due to the large error bars, the scatter in the full set of residuals is less than 5 minutes. We see no evidence for long term drifts in the period.

We used six of the eight EPOCh secondary eclipses of TrES-2 to place a 95\% confidence upper limit on the eclipse depth of 0.16\%. This corresponds to a planetary geometric albedo of $A_g = 6.6$. As for HAT-P-4, this is not a physically plausible value.

\subsection{WASP-3}
\label{sec:wasp3} 

\begin{figure}[h!]
  \begin{center}
    \includegraphics[width=4in, angle=90]{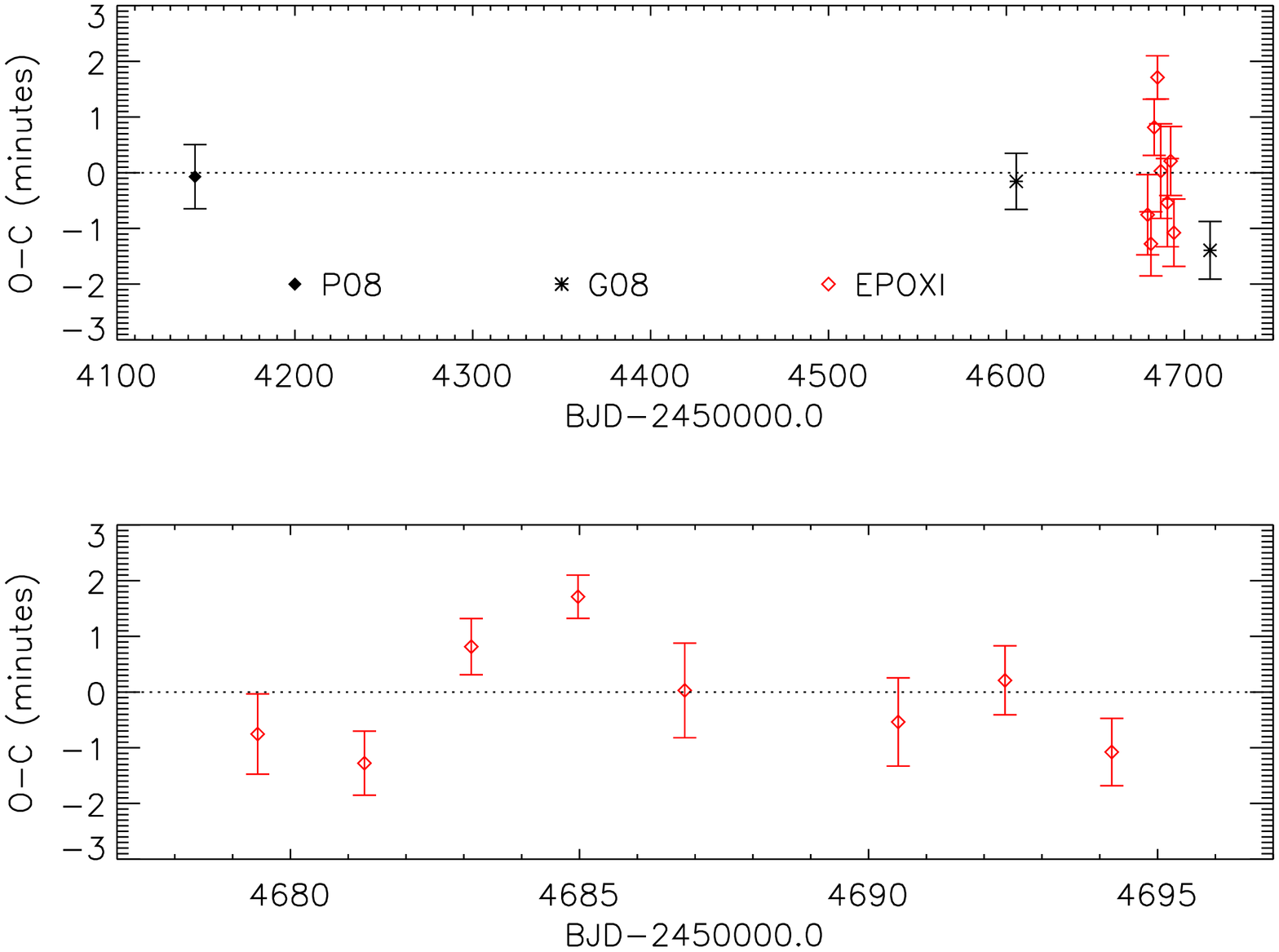}
    \caption{The transit times of WASP-3. P08: \citet{Pollacco08}; G08: \citet{Gibson08}; EPOXI: this paper. {\it Lower panel}: An expanded view of the EPOXI transit times, with $1\sigma$ errors of 23--51 seconds.}
    \label{fig:wasp3_ttvs}
  \end{center}
\end{figure}

For WASP-3, we measure system parameters that are consistent with, and an improvement upon, previously published parameters from \citet{Pollacco08} and \citet{Gibson08}, finding $R_p=1.385\pm0.060$ $R_{\rm Jup}$, $R_{\star}=1.354\pm0.056$ $R_{\odot}$, $i=84.22\pm0.81$ degrees and $\tau=167.3\pm1.3$ minutes. We generate a new refined ephemeris from the published transit times and the eight EPOCh transits in this paper, finding $T_c({\rm BJD})=2454686.82069\pm0.00039+1.8468373\pm0.0000014E$. The residuals to this ephemeris are shown in Figure \ref{fig:wasp3_ttvs}.

\begin{figure}[h!]
  \begin{center}
    \includegraphics[width=4in, angle=90]{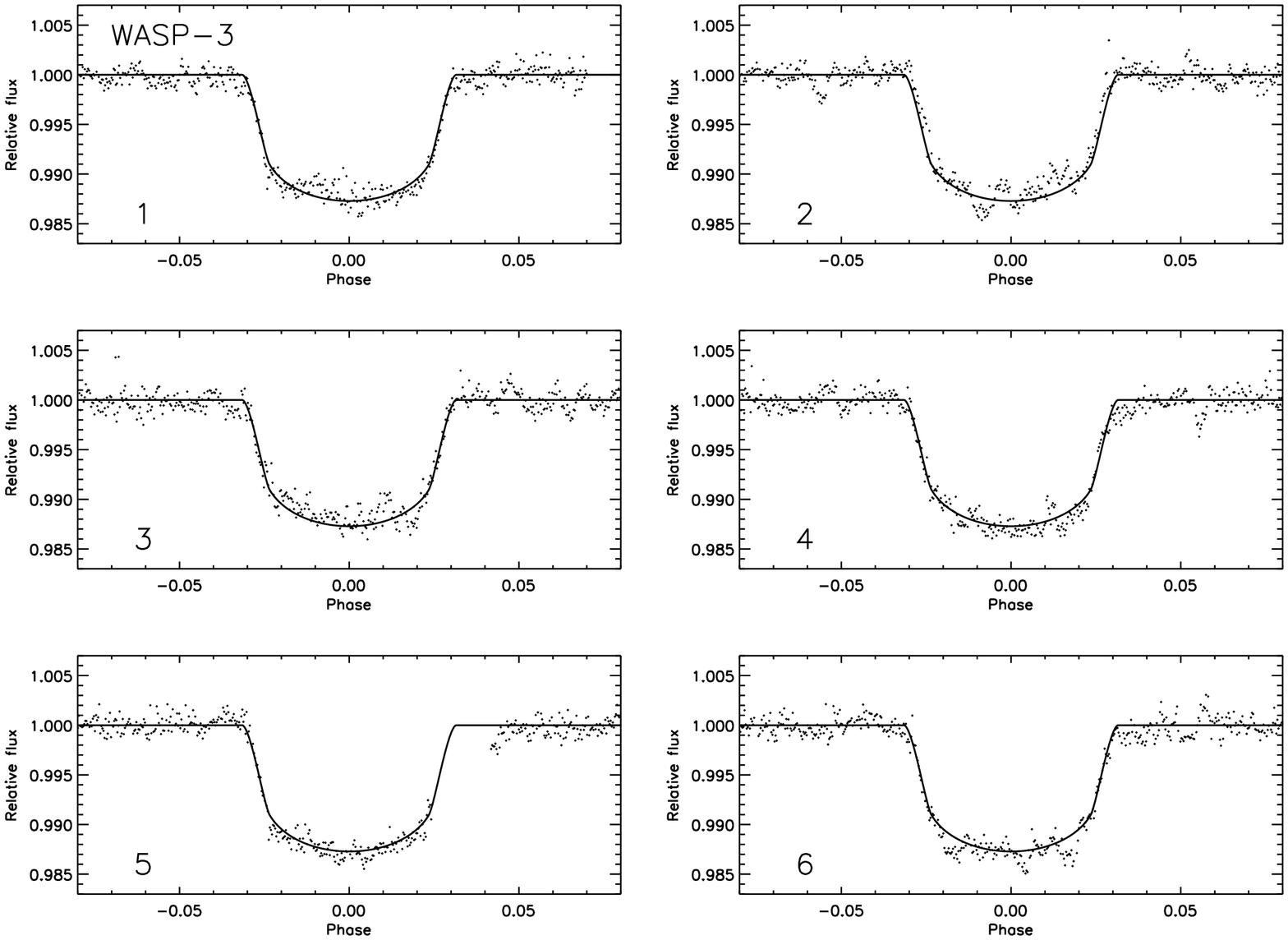}
    \caption{The six EPOCh transits of WASP-3. The best fit model for the combined set of transits is plotted in each case. The scatter around the model is not typically larger in transit than out of transit, indicating that the in-transit residuals cannot be attributed to star spots.}
    \label{fig:wasp3_transits}
  \end{center}
\end{figure}

The phase-folded light curve of WASP-3 (Figure \ref{fig:wasp3_phased}) shows correlated residuals in the latter half of transit. Since the noise in the transit exceeds the noise out of transit, one conclusion could be spot activity on the surface of the star being eclipsed during transit. However, if we examine the transits individually we observe that the correlated noise in the full light curve is not typically larger in transit than out of transit. The six transits used in the analysis are shown in Figure \ref{fig:wasp3_transits}. In the transits numbered 2, 3, 4, and 6 large deviations can be seen in the second half of the transit, which leads to residuals in the phased light curve. If there were star spots producing correlated residuals in the transits, we would not necessarily expect them to occur at the same phase for each transit. The $v$ sin $i$ for WASP-3 has been measured by \citet{Simpson09} to be $15.7^{+1.4}_{-1.3}$ km s$^{-1}$, which corresponds to a rotational period for the star of 4.2 days. Transits of WASP-3 are spaced by 1.85 days, so it is improbable for spot activity to appear at the same phase in successive transits. Given these constraints, we conclude that the alignment of signals with phase in the EPOCh transits of WASP-3 are coincidental and are due to instrumental artifacts.

We use four of the nine EPOCh secondary eclipse observations of WASP-3 to set a 95\% confidence upper limit on the eclipse depth of 0.11\%. This corresponds to a planetary geometric albedo of $A_g=2.5$ in the EPOCh bandpass, which is not a useful constraint for the planetary atmosphere.

\section{Conclusion}
\label{sec:conc}

We have presented time series photometry from the NASA {\it EPOXI} Mission of Opportunity for four known transiting planet systems: HAT-P-4, TrES-3, TrES-2 and WASP-3. For each system we provided an updated set of system parameters and orbital period, and placed upper limits on the secondary eclipse depth. For TrES-3, we see evidence of stellar variability over long timescales. We combined the {\it EPOCh} secondary eclipse upper limit for TrES-3 with previously published measurements and confirm that the data are best fit using an atmosphere model with no temperature inversion. For TrES-2, the EPOCh data weaken the claimed trends of decreasing inclination and transit duration \citep{Mislis09,Mislis10}. We have also performed a search for additional transiting planets in the EPOCh photometry for these systems, which we will present in a forthcoming paper.

\acknowledgments

We are extremely grateful to the {\it EPOXI}~Flight and Spacecraft Teams that made these difficult observations possible.  At the Jet Propulsion Laboratory, the Flight Team has included M. Abrahamson, B. Abu-Ata, A.-R. Behrozi, S. Bhaskaran, W. Blume, M. Carmichael, S. Collins, J. Diehl, T. Duxbury, K. Ellers, J. Fleener, K. Fong, A. Hewitt, D. Isla, J. Jai, B. Kennedy, K. Klassen, G. LaBorde, T. Larson, Y. Lee, T. Lungu, N. Mainland, E. Martinez, L. Montanez, P. Morgan, R. Mukai, A. Nakata, J. Neelon, W. Owen, J. Pinner, G. Razo Jr., R. Rieber, K. Rockwell, A. Romero, B. Semenov, R. Sharrow, B. Smith, R. Smith, L. Su, P. Tay, J. Taylor, R. Torres, B. Toyoshima, H. Uffelman, G. Vernon, T. Wahl, V. Wang, S. Waydo, R. Wing, S. Wissler, G. Yang, K. Yetter, and S. Zadourian.  At Ball Aerospace, the Spacecraft Systems Team has included L. Andreozzi, T. Bank, T. Golden, H. Hallowell, M. Huisjen, R. Lapthorne, T. Quigley, T. Ryan, C. Schira, E. Sholes, J. Valdez, and A. Walsh. 

Support for this work was provided by the {\it EPOXI} Project of the National Aeronautics and Space Administration's Discovery Program via funding to the Goddard Space Flight Center, and to Harvard University via Co-operative Agreement NNX08AB64A, and to the Smithsonian Astrophysical Observatory via Co-operative Agreement NNX08AD05A. The authors acknowledge and are grateful for the use of publicly available transit modeling routines by Eric Agol and Kaisey Mandel, and also the Levenberg-Marquardt least-squares minimization routine MPFITFUN by Craig Markwardt. This work has used data obtained by various observers collect in the Exoplanet Transit Database, http://var.astro.cz/ETD.

\newpage

\clearpage

\end{document}